\documentclass[pra,showpacs,showkeys]{revtex4}

\usepackage{bm}
\usepackage{graphicx}
\usepackage{amssymb}
\usepackage{amsmath}
\usepackage{amsthm}
\usepackage{mathrsfs}

\begin{document}

\title{Probabilities of failure for quantum error correction}

\author{A. J. Scott}
\email{ascott@phys.unm.edu} \affiliation{Department of Physics
and Astronomy, University of New Mexico, Albuquerque, NM 87131-1156,
USA}

\begin{abstract}
We investigate the performance of a quantum error-correcting code when pushed beyond its intended 
capacity to protect information against errors, presenting formulae for the probability of failure when the errors affect more qudits 
than that specified by the code's minimum distance. Such formulae provide a means to rank different 
codes of the same minimum distance. We consider both error detection and error correction, treating 
explicit examples in the case of stabilizer codes constructed from qubits and encoding a single qubit.
\end{abstract}

\maketitle

\def\w{\omega}
\def\wb{\overline{\omega}}
\def\tr{\operatorname{tr}}
\def\Tr{\operatorname{Tr}}
\def\wt{\operatorname{wt}}
\def\supp{\operatorname{supp}}
\def\prob{\operatorname{Prob}}
\def\ket#1{|#1\rangle}
\def\bra#1{\langle#1|}
\def\inner#1#2{\langle #1 | #2 \rangle}
\def\Exp{\mathop{\operatorname{E}}}
\def\Var{\mathop{\operatorname{Var}}}

\def\T{\mathcal{T}^{\scriptscriptstyle d}}
\def\Fd{\mathcal{F}^{\scriptscriptstyle d}}
\def\Fc{\mathcal{F}^{\scriptscriptstyle c}}
\def\fd{\mathscr{F}^{\scriptscriptstyle d}}
\def\fc{\mathscr{F}^{\scriptscriptstyle c}}

\theoremstyle{plain}
\newtheorem{thm}{Theorem}
\newtheorem{lem}[thm]{Lemma}
\newtheorem{cor}[thm]{Corollary}
\newtheorem{prp}[thm]{Proposition}

\theoremstyle{definition}
\newtheorem{dfn}[thm]{Definition}

\theoremstyle{remark}
\newtheorem*{rmk}{Remark}
\newtheorem*{pprf}{Proof}
\newtheorem{exm}{Example}

\newenvironment{prf}{\begin{pprf}}{$\Box$\end{pprf}}

\setcounter{topnumber}{1}

\section{Introduction}
\label{sec1}

Quantum error-correcting codes~\cite{calderbank,gottesman,knill,preskill,nielsen,grassl} protect quantum information 
against noise. They play important roles in many areas of quantum information theory, but most critically, in the 
viability of a quantum computer. Quantum error correction negates a quantum state's natural susceptibility 
to decohere, and thus provides the long-time coherence necessary to sustain quantum computation. Shor~\cite{shor2} 
and Steane~\cite{steane} presented the first constructions of quantum error-correcting codes. These discoveries led to a 
formal connection between quantum codes and classical additive codes~\cite{calderbank}, and consequently, the 
characterization of a general class of quantum error-correcting codes commonly referred to as stabilizer codes~\cite{gottesman}.

The idea behind quantum error correction is to encode quantum states into qudits in such a way that a small number of 
errors affecting the individual qudits can be detected and corrected to perfectly restore the original encoded state.
In this article we investigate the integrity of a quantum error-correcting code when under the influence of errors affecting 
more qudits than what the code was originally designed to handle. We derive general formulae to calculate the probabilities 
of successful error detection or correction, when the errors are depolarizing, and we are given either (i) the  
location of the errors, (ii) the number of errors, or, (iii) the probability that a single qudit is in error. 
Such formulae provide a means to compare codes of the same minimum distance. For the analysis of error detection we treat 
general quantum error-correcting codes constructed from qudits. This extends the results of Ashikhmin {\it et al}~\cite{ashikhmin2}. We 
then specialize to stabilizer codes for the case of error correction, where the dimension of the constituent qudit 
subsystems is a prime power.

When the constituent subsystems are qubits, we give explicit results for a variety of stabilizer codes encoding a single qubit.  
We find that, for the depolarizing channel under error detection, as the number of qubits increases we are generally able to 
construct better codes even when the minimum distance remains constant. However this is not so for error correction. In this case
the unique five-qubit quantum Hamming code outperforms all other codes of minimum distance three.

The article is organized as follows. In the next section we introduce quantum error-correcting codes, giving conditions for a code   
to have a specified minimum distance in terms of its weight enumerators. We then build on this treatment to analyze error detection 
in Sec.~\ref{sec3}, and error correction in Sec.~\ref{sec4}, where stabilizer codes are also introduced. In Sec.~\ref{sec5} we 
characterize stabilizer codes constructed from qubits in terms of classical additive codes, giving many examples which we then 
investigate. Finally, in Sec.~\ref{sec6} we review the main results of the article. 

\section{Quantum error-correcting codes}
\label{sec2}

The idea behind quantum error correction \cite{calderbank,gottesman,knill,preskill,nielsen,grassl} 
is to encode quantum states into qudits in such a way that a small number of errors affecting the individual 
qudits can be measured and corrected to perfectly restore the original encoded state. The encoding of a 
$K$-dimensional quantum state into $n$ qudits is simply a linear map from $\mathbb{C}^K$ to a subspace 
$\mathcal{Q}$ of $(\mathbb{C}^D)^{\otimes n}$. The subspace itself is referred to as the {\it code} and is 
orientated in such a way that errors on the qudits move encoded states in a direction perpendicular to the code.
We will refer to such codes as $((n,K))_D$ quantum error-correcting codes.

An {\it error operator} $E$ is a linear operator acting on $(\mathbb{C}^D)^{\otimes n}$. The error 
is said to be {\it detectable} by the quantum code $\mathcal{Q}$ if 
\begin{equation}
\bra{\psi}E\ket{\psi}=\bra{\phi}E\ket{\phi}
\end{equation}
for all normalized $\ket{\psi},\ket{\phi}\in\mathcal{Q}$. It is a general theorem of quantum error correction that a set of 
errors $\mathcal{E}$ can be {\it corrected} by a code $\mathcal{Q}$, if and only if for each 
$E_1,E_2\in\mathcal{E}$, the error $E_2^\dag E_1$ is detectable by $\mathcal{Q}$ \cite{knill}. 

Define the {\it support} of an error operator $E$, denoted by $\supp(E)$, as the subset of $\{1,\dots,n\}$ 
consisting of all indices labeling a qudit where $E$ acts nontrivially i.e. $E$ is not a scalar 
multiple of the identity on the qudit. The {\it weight} of $E$ is then the cardinality of $\supp(E)$,
$\wt(E)\equiv |\supp(E)|$. A quantum code $\mathcal{Q}$ has a {\it minimum distance} of at least $d$ if and 
only if all errors of weight less than $d$ are detectable by $\mathcal{Q}$. A code with minimum distance $d$ 
allows the correction of arbitrary errors affecting $<d/2$ qudits. Such codes are denoted by the triple 
$((n,K,d))_D$. An $((n,K,d))_D$ code is called {\it pure} if $\bra{\psi}E\ket{\psi}=D^{-n}\tr{E}$ for all 
$\ket{\psi}\in\mathcal{Q}$ whenever $\wt(E)<d$. The notion of pure is equivalent to {\it nondegenerate} for 
stabilizer codes \cite{calderbank,gottesman}. When considering {\it self-dual} codes ($K=1$), we adopt the convention 
that the notation $((n,1,d))_D$ refers only to pure codes since the condition on the minimum distance is otherwise trivial.

The remainder of this section is based on an article on quantum weight enumerators by Rains \cite{rains4}. We start with a lemma.

\begin{lem}\label{lemsym}
Let $\mathcal{Q}\leq\mathbb{C}^N$ with dimension $K$ and associated projector $P$. Furthermore denote by $\Exp_{\psi\in\mathcal{Q}}[\,\cdot\,]$ the 
unitarily invariant uniform average over all $\ket{\psi}\in\mathcal{Q}$. Then
\begin{equation}
\Exp_{\psi\in\mathcal{Q}}\left[\ket{\psi}\bra{\psi}^{\otimes t}\right]=\frac{t!(K-1)!}{(t+K-1)!}\Pi_\text{\rm sym}^t \label{symavg}
\end{equation}
where $\Pi_\text{\rm sym}^t$ is the projector onto the totally symmetric subspace of $\mathcal{Q}^{\otimes t}$. In particular
\begin{equation}
\Exp_{\psi\in\mathcal{Q}}\left[\ket{\psi}\bra{\psi}\right]=\frac{P}{K} \qquad\text{and}\qquad \Exp_{\psi\in\mathcal{Q}}\left[\ket{\psi}\bra{\psi}\otimes\ket{\psi}\bra{\psi}\right]=\frac{(P\otimes P)(1+T)}{K(K+1)} 
\end{equation}
where $T$ is the swap on $\mathbb{C}^N\!\!\otimes\mathbb{C}^N$ i.e. $T\ket{\psi}\otimes\ket{\phi}=\ket{\phi}\otimes\ket{\psi}$ for all 
$\ket{\psi}, \ket{\phi}\in\mathbb{C}^N$. 
\end{lem}
\begin{prf}
Use Schur's lemma. Eq. (\ref{symavg}) is invariant under all unitaries $U^{\otimes t}$ which act irreducibly on the totally symmetric subspace 
$\Pi_\text{sym}^t\mathcal{Q}^{\otimes t}$. Also note that $\Pi_\text{sym}^1=P$ and $\Pi_\text{sym}^2=(P\otimes P)(1+T)/2$. 
\end{prf}

Consider the variance in $\bra{\psi}E\ket{\psi}$ over all states in the code
\begin{equation}
\Var_{\psi\in\mathcal{Q}}\left[\bra{\psi}E\ket{\psi}\right]\equiv\Exp_{\psi\in\mathcal{Q}}\left[\big|\bra{\psi}E\ket{\psi}-\Exp_{\phi\in\mathcal{Q}}\left[\bra{\phi}E\ket{\phi}\right]\big|^2\right]. \label{var}
\end{equation}
Using Lemma \ref{lemsym} we have the mean 
\begin{equation}
\Exp_{\psi\in\mathcal{Q}}\left[\bra{\psi}E\ket{\psi}\right]=\tr\left(E\Exp_{\psi\in\mathcal{Q}}\left[\ket{\psi}\bra{\psi}\right]\right)=\frac{1}{K}\tr\left(EP\right)
\end{equation}
and second moment
\begin{eqnarray}
\Exp_{\psi\in\mathcal{Q}}\left[\bra{\psi}E^\dag\ket{\psi}\bra{\psi}E\ket{\psi}\right] &=& \tr\left\{\left(E^\dag\otimes E\right) \Exp_{\psi\in\mathcal{Q}}\left[\ket{\psi}\bra{\psi}\otimes\ket{\psi}\bra{\psi}\right]\right\} \\
&=&\frac{1}{K(K+1)}\tr\left[\left(E^\dag\otimes E\right)\left(P\otimes P\right)(1+T)\right] \\
&=&\frac{1}{K(K+1)}\left[\tr\left(E^\dag PEP\right)+\tr\left(E^\dag P\right)\tr\left(EP\right)\right] \label{misceq1}
\end{eqnarray}
where we have used the fact that $\tr[(A\otimes B)T]=\tr(AB)$, and thus
\begin{eqnarray}
\Var_{\psi\in\mathcal{Q}}\left[\bra{\psi}E\ket{\psi}\right] &=& \Exp_{\psi\in\mathcal{Q}}\left[\bra{\psi}E^\dag\ket{\psi}\bra{\psi}E\ket{\psi}\right] - \big|\Exp_{\psi\in\mathcal{Q}}\left[\bra{\psi}E\ket{\psi}\right]\big|^2 \\
&=& \frac{1}{K^2(K+1)}\left[K\tr\left(E^\dag PEP\right)-\tr\left(E^\dag P\right)\tr\left(EP\right)\right].
\end{eqnarray}
Now, by noting that an error $E$ is detectable if and only if the variance vanishes, we have 
the following equivalent definition of error detection.

\begin{lem}
Let $\mathcal{Q}$ be an $((n,K))_D$ quantum code with associated projector $P$. Then the error $E$ is 
detectable by $\mathcal{Q}$ iff $K\tr\left(E^\dag PEP\right)=\tr\left(E^\dag P\right)\tr\left(EP\right)$.
\end{lem}

The multi-qudit displacement operators 
\begin{equation}
\mathcal{D}(\bm{\mu},\bm{\nu})\equiv\mathcal{D}(\mu_1\dots\mu_n,\nu_1\dots\nu_n)\equiv D(\mu_1,\nu_1)\otimes\cdots\otimes D(\mu_n,\nu_n)\qquad 0\leq\mu_k,\nu_k\leq D-1
\end{equation}
where
\begin{equation}
D(\mu,\nu)\equiv e^{i\pi\mu\nu/D}X^{\mu}Z^{\nu}, \qquad X\ket{j}\equiv\ket{j+1 \text{ mod } D}, \qquad Z\ket{j}\equiv e^{2\pi ij/D}\ket{j},
\end{equation}
form an orthonormal basis for the set of all $n$-qudit operators: 
\begin{equation}
A=D^{-n}\sum_{\bm{\mu},\bm{\nu}}\tr[\mathcal{D}(\bm{\mu},\bm{\nu})^\dag A]\mathcal{D}(\bm{\mu},\bm{\nu}).
\end{equation} 
The weight of $\mathcal{D}(\bm{\mu},\bm{\nu})$ is simply the number of pairs $(\mu_k,\nu_k)$ different from 
$(0,0)$. For future reference we now note some properties of displacement operators:
\begin{eqnarray}
D(\mu,\nu) &=& e^{i\pi\nu}D(\mu+D,\nu) \;=\; e^{i\pi\mu}D(\mu,\nu+D) \label{disp1}\\
D(\mu,\nu)^\dag &=& D(-\mu,-\nu) \;=\; e^{i\pi(\mu+\nu+D)}D(D-\mu,D-\nu) \label{disp2}\\
D(\mu,\nu)^D &=& D(\mu D,\nu D) \;=\; e^{i\pi\mu\nu D}I \label{disp3}\\
D(\mu,\nu)D(\alpha,\beta) &=& e^{2\pi i(\nu\alpha-\mu\beta)/D}D(\alpha,\beta)D(\mu,\nu) \;=\; e^{\pi i(\nu\alpha-\mu\beta)/D}D(\mu+\alpha,\nu+\beta) \label{disp4}\\
\tr\left[D(\mu,\nu)^{\dag}D(\alpha,\beta)\right] &=& D\delta_{\mu\alpha}\delta_{\nu\beta}.\label{disp5}
\end{eqnarray}

Note that if the errors $E_1$ and $E_2$ are detectable, then any linear combination $c_1E_1+c_2E_2$ 
is also detectable. In particular, a linear space of errors $\left\{\sum c_i E_i | c_i\in\mathbb{C}\right\}$ is 
detectable if and only if all errors $E_i$ are detectable. By defining the enumerators 
\begin{eqnarray}
A_S'(P) &\equiv& \frac{D^{|S|}}{K^2}\Exp_{\supp\mathcal{D}(\bm{\mu},\bm{\nu})\subseteq S}\left[\tr\left\{\mathcal{D}(\bm{\mu},\bm{\nu})^\dag P\right\} \tr\left\{\mathcal{D}(\bm{\mu},\bm{\nu})P\right\}\right] \label{AS2}\\
&\equiv& \frac{1}{D^{|S|}K^2}\sum_{\supp \mathcal{D}(\bm{\mu},\bm{\nu})\subseteq S} \tr\left[\mathcal{D}(\bm{\mu},\bm{\nu})^\dag P\right] \tr\left[\mathcal{D}(\bm{\mu},\bm{\nu})P\right] \label{AS3}\\
B_S'(P)&\equiv& \frac{D^{|S|}}{K}\Exp_{\supp\mathcal{D}(\bm{\mu},\bm{\nu})\subseteq S}\left[\tr\left\{\mathcal{D}(\bm{\mu},\bm{\nu})^\dag P\mathcal{D}(\bm{\mu},\bm{\nu})P\right\}\right] \label{BS2}\\
&\equiv&\frac{1}{D^{|S|}K}\sum_{\supp \mathcal{D}(\bm{\mu},\bm{\nu})\subseteq S} \tr\left[\mathcal{D}(\bm{\mu},\bm{\nu})^\dag P\mathcal{D}(\bm{\mu},\bm{\nu})P\right] \label{BS3}
\end{eqnarray}
and noting that $D^{|S|}\left[B_S'(P)-A_S'(P)\right]/(K+1)$ is the sum over all positive quantities 
$\Var_{\psi\in\mathcal{Q}}\left[\bra{\psi}\mathcal{D}(\bm{\mu},\bm{\nu})\ket{\psi}\right]$ with 
$\supp \mathcal{D}(\bm{\mu},\bm{\nu})\subseteq S$, we have the following lemma.

\begin{lem}\label{lemdetectS}
Let $\mathcal{Q}$ be an $((n,K))_D$ quantum code with associated projector $P$. Then all errors $E$ with 
$\supp E\subseteq S$ are detectable by $\mathcal{Q}$ iff $B_S'(P)=A_S'(P)$. 
\end{lem}

With the help of Eq. (\ref{disp4}) we find that
\begin{eqnarray}
\Exp_{\bm{\mu},\bm{\nu}}\left[\mathcal{D}(\bm{\mu},\bm{\nu})^\dag A\mathcal{D}(\bm{\mu},\bm{\nu})\right] 
&\equiv& D^{-2n}\sum_{\bm{\mu},\bm{\nu}}\mathcal{D}(\bm{\mu},\bm{\nu})^\dag A\mathcal{D}(\bm{\mu},\bm{\nu}) \\
&=& D^{-3n}\sum_{\bm{\mu},\bm{\nu},\bm{\alpha},\bm{\beta}}\tr[\mathcal{D}(\bm{\alpha},\bm{\beta})^\dag A]\mathcal{D}(\bm{\mu},\bm{\nu})^\dag\mathcal{D}(\bm{\alpha},\bm{\beta}) \mathcal{D}(\bm{\mu},\bm{\nu}) \\
&=& D^{-3n}\sum_{\bm{\mu},\bm{\nu},\bm{\alpha},\bm{\beta}}\tr[\mathcal{D}(\bm{\alpha},\bm{\beta})^\dag A]e^{2\pi i(\bm{\mu}\cdot\bm{\beta}-\bm{\alpha}\cdot\bm{\nu})/D}\mathcal{D}(\bm{\alpha},\bm{\beta}) \\
&=& D^{-n}\sum_{\bm{\alpha},\bm{\beta}}\tr[\mathcal{D}(\bm{\alpha},\bm{\beta})^\dag A]\delta_{\bm{\alpha}\bm{0}}\delta_{\bm{\beta}\bm{0}}\mathcal{D}(\bm{\alpha},\bm{\beta}) \\
&=& D^{-n}\tr(A) I. \label{displacerand}
\end{eqnarray} 
The following lemma now applies to the operators $\mathcal{D}(\bm{\mu},\bm{\nu})$.

\begin{lem}\label{lemUps}
Let $\Exp_{\Upsilon}[\,\cdot\,]$ denote the expectation (average), given some probability measure $d\mu(\Upsilon)$, over a set of linear operators $\{\Upsilon\}\subset\text{\rm End}(\mathbb{C}^N)$. Then the following three statements are equivalent.
\begin{enumerate}
\item $\Exp_{\Upsilon}\left[\tr(\Upsilon^\dag A) \Upsilon \right]=A/N$ for all linear operators $A$,
\item $\Exp_{\Upsilon}\left[\Upsilon^\dag A \Upsilon \right]=\tr(A)I/N$ for all linear operators $A$,
\item $\Exp_{\Upsilon}\left[\Upsilon^\dag \otimes \Upsilon \right]=T/N$, where $T$ is the swap.
\end{enumerate}
\end{lem}
\begin{prf}
Since only 2 $\Leftrightarrow$ 3 is needed for the current article we will leave the remaining parts of the 
proof as an exercise for the reader. 

Assuming 2 and letting $\ket{j}$ be a basis for $\mathbb{C}^N$, we have 
\begin{equation}
\bra{j}\otimes\bra{k}\left({\textstyle\Exp_{\Upsilon}}\left[\Upsilon^\dag\otimes\Upsilon\right]\right)\ket{l}\otimes\ket{m}
=\bra{j}{\textstyle\Exp_{\Upsilon}}\left[\Upsilon^\dag\ket{l}\bra{k}\Upsilon\right]\ket{m}=\delta_{kl}\delta_{jm}/N
\end{equation}
and thus 2 $\Rightarrow$ 3. Assuming 3, 
\begin{eqnarray}
\bra{l}{\textstyle\Exp_{\Upsilon}}\left[\Upsilon^\dag A\Upsilon\right]\ket{m} &=& \sum_{j,k}{\textstyle\Exp_{\Upsilon}}\left[\bra{l}\Upsilon^\dag\ket{j}\bra{j}A\ket{k}\bra{k}\Upsilon\ket{m}\right] \\
&=&\sum_{j,k}\bra{l}\otimes\bra{k}\left({\textstyle\Exp_{\Upsilon}}\left[\Upsilon^\dag\otimes\Upsilon\right]\right)\ket{j}\otimes\ket{m}\bra{j}A\ket{k}\\
&=&\sum_{j,k}\delta_{jk}\delta_{ml}\bra{j}A\ket{k}/N=\tr(A)\delta_{lm}/N
\end{eqnarray}
and thus 3 $\Rightarrow$ 2. 
\end{prf}

With Lemma \ref{lemUps} in hand we find that
\begin{eqnarray}
\Exp_{\supp\mathcal{D}(\bm{\mu},\bm{\nu})\subseteq S}\left[\tr\left\{\mathcal{D}(\bm{\mu},\bm{\nu})^\dag P\right\}\tr\left\{\mathcal{D}(\bm{\mu},\bm{\nu})P\right\}\right] &=& \tr\left\{\Exp_{\supp\mathcal{D}(\bm{\mu},\bm{\nu})\subseteq S}\left[\mathcal{D}(\bm{\mu},\bm{\nu})^\dag\otimes \mathcal{D}(\bm{\mu},\bm{\nu})\right](P\otimes P)\right\} \\
&=& D^{-|S|}\tr\left[T_S(P\otimes P)\right] \\
&=& D^{-|S|}\tr_S\left[(\tr_{S'}P)^2\right]
\end{eqnarray}
where $T_S$ acts on $(\mathbb{C}^D)^{\otimes n}\otimes(\mathbb{C}^D)^{\otimes n}$ by swapping all qudits with 
indices in $S$, and
\begin{eqnarray}
\Exp_{\supp\mathcal{D}(\bm{\mu},\bm{\nu})\subseteq S}\left[\tr\left\{\mathcal{D}(\bm{\mu},\bm{\nu})^\dag P\mathcal{D}(\bm{\mu},\bm{\nu})P\right\}\right] 
&=& D^{-|S|}\tr\left[(\tr_SP)P\right] \\
&=& D^{-|S|}\tr_{S'}\left[(\tr_SP)^2\right].
\end{eqnarray}
Consequently, our previously cumbersome definition of the weight enumerators (\ref{AS2}-\ref{BS3}) may be simplified to
\begin{eqnarray}
A_S'(P) &=& \frac{1}{K^2}\tr_S\left[(\tr_{S'}P)^2\right] \label{AS4}\\
B_S'(P) &=& \frac{1}{K}\tr_{S'}\left[(\tr_SP)^2\right]. \label{BS4}
\end{eqnarray}
It is easily verified that the normalization condition
$A_\emptyset'(P)=B_\emptyset'(P)=1$ is satisfied, for self-dual codes 
$B_S'(P)=A_S'(P)$ for all $S\subseteq\{1,\dots,n\}$, and in general,  
\begin{equation}
B_S'(P)\geq A_S'(P)>0 
\end{equation}
for all $S\subseteq\{1,\dots,n\}$. Also note that the above simplification (\ref{AS4},\ref{BS4}) gives the relation 
\begin{equation}
B_S'(P)=KA_{S'}'(P). \label{relS}
\end{equation} 
Note from Eq. (\ref{AS3}) that $D^{|T|}A_T'(P)\leq D^{|S|}A_S'(P)$ whenever $T\subseteq S$. We obtain 
a similar inequality from Eq. (\ref{BS3}), and with the above relation (\ref{relS}), one may deduce 
that 
\begin{eqnarray}
D^{|T|-|S|}A_T'(P)\leq \!&A_S'(P)&\!\leq D^{|S|-|T|}A_T'(P) \label{rankA}\\
D^{|T|-|S|}B_T'(P)\leq \!&B_S'(P)&\!\leq D^{|S|-|T|}B_T'(P) \label{rankB}
\end{eqnarray}
whenever $T\subseteq S$. By diagonalizing $P=\sum_{k=1}^K \ket{k}\bra{k}$, and using the Cauchy-Schwarz 
inequality, $|\tr AB^\dag|^2\leq\tr AA^\dag\tr BB^\dag$, we find that 
$\tr_S\left[(\tr_{S'}P)^2\right] = \sum_{j,k=1}^K\tr_S(\rho_j\rho_k)\leq \sum_{j,k=1}^K\sqrt{\tr_S({\rho_j}^2)\tr_S({\rho_k}^2)}\leq K^2$
since $\tr_S({\rho_k}^2)\leq 1$ where $\rho_k\equiv\tr_{S'}\ket{k}\bra{k}$. Now using this inequality on 
Eq.'s (\ref{AS4}) and (\ref{BS4}), and with $T=\emptyset$ in inequalities (\ref{rankA}) and (\ref{rankB}) we obtain various 
bounds for the weight enumerators. These reduce to
\begin{eqnarray}
\max\left\{D^{-|S|},D^{|S|-n}/K\right\}\leq \!&A_S'(P)&\!\leq\min\left\{1,D^{n-|S|}/K\right\} \label{boundAS}\\
\max\left\{D^{-|S|},KD^{|S|-n}\right\}\leq \!&B_S'(P)&\!\leq\min\left\{D^{|S|},K\right\}. \label{boundBS}
\end{eqnarray}  
Finally we remark that the weight enumerators are unchanged if we replace the operators 
$\mathcal{D}(\bm{\mu},\bm{\nu})$ used in the expectations (\ref{AS2}) and (\ref{BS2}) by 
any type of random depolarizing error (Definition \ref{depolar}). In particular we 
could choose random unitaries under the Haar measure as was done in their original definition \cite{rains4}.

We now define the Rains enumerators \cite{rains4} 
\begin{eqnarray}
A_m'(P)&\equiv&\sum_{|S|=m}A_S'(P)=\frac{1}{K^2}\sum_{|S|=m}\tr_S\left[(\tr_{S'}P)^2\right] \label{Amp}\\
B_m'(P)&\equiv&\sum_{|S|=m}B_S'(P)=\frac{1}{K}\sum_{|S|=m}\tr_{S'}\left[(\tr_SP)^2\right] \label{Bmp}
\end{eqnarray}
where $m=0,\dots,n$. These satisfy the normalization condition
$A_0'(P)=B_0'(P)=1$, for self-dual codes $B_m'(P)=A_m'(P)$ for all $0\leq m\leq n$, 
and in general, 
\begin{equation}
B_m'(P)\geq A_m'(P)>0
\end{equation}
for all $0\leq m\leq n$. Again we have the relation 
\begin{equation}
B_m'(P)=KA_{n-m}'(P) \label{rel3}
\end{equation} 
and the inequalities
\begin{eqnarray}
D^{l-m}A_l'(P)\leq \!&A_m'(P)&\!\leq D^{m-l}A_l'(P) \\
D^{l-m}B_l'(P)\leq \!&B_m'(P)&\!\leq D^{m-l}B_l'(P)
\end{eqnarray}
whenever $l\leq m$, and
\begin{eqnarray}
\max\left\{D^{-m},D^{m-n}/K\right\}\leq \!& \displaystyle\frac{m!(n-m)!}{n!}A_m'(P)&\!\leq\min\left\{1,D^{n-m}/K\right\} \label{boundAm}\\
\max\left\{D^{-m},KD^{m-n}\right\}\leq \!& \displaystyle\frac{m!(n-m)!}{n!}B_m'(P)&\!\leq\min\left\{D^{m},K\right\}. \label{boundBm}
\end{eqnarray}  

Finally, by noticing that $B_m'(P)-A_m'(P)=\sum_{|S|=m}B_S'(P)-A_S'(P)=0$ if and only if each term in the sum is zero, 
as a simple consequence of Lemma \ref{lemdetectS}, we have the following theorem \cite{rains4}.

\begin{thm}\label{thmRains}
Let $\mathcal{Q}$ be an $((n,K))_D$ quantum code with associated 
projector $P$. Then $\mathcal{Q}$ has minimum distance of at least $d$ iff $B_{d-1}'(P)=A_{d-1}'(P)$. 
\end{thm}

Alternatively, one may define the Shor-Laflamme enumerators \cite{shor}
\begin{eqnarray}
A_m(P)&\equiv&\frac{1}{K^2}\sum_{\wt\mathcal{D}(\bm{\mu},\bm{\nu})=m} \tr[\mathcal{D}(\bm{\mu},\bm{\nu})^\dag P] \tr[\mathcal{D}(\bm{\mu},\bm{\nu})P] \label{Ashor}\\
B_m(P)&\equiv&\frac{1}{K}\sum_{\wt\mathcal{D}(\bm{\mu},\bm{\nu})=m} \tr[\mathcal{D}(\bm{\mu},\bm{\nu})^\dag P\mathcal{D}(\bm{\mu},\bm{\nu})P] \label{Bshor}
\end{eqnarray}
where $m=0,\dots,n$, which [from Eq.'s (\ref{AS3}) and (\ref{BS3})] may be related to the Rains enumerators   
\begin{eqnarray}
A_m'(P)&=& D^{-m} \sum_{i=0}^m \frac{(n-i)!}{(m-i)!(n-m)!}A_i(P) \label{rel1}\\
B_m'(P)&=& D^{-m} \sum_{i=0}^m \frac{(n-i)!}{(m-i)!(n-m)!}B_i(P) \label{rel2}
\end{eqnarray}
and satisfy the normalization condition $A_0(P)=B_0(P)=1$, for self-dual codes $B_i(P)=A_i(P)$ 
for all $0\leq i\leq n$, and in general $B_i(P)\geq A_i(P)\geq 0$ for $0\leq i\leq n$. 
The above relations [Eq.'s (\ref{rel3}), (\ref{rel1}) and (\ref{rel2})] may be used to derive quantum versions of the MacWilliams identities, 
and thus, bounds on the parameters of a quantum code \cite{shor,rains4}. Such bounds are the principle reason for defining quantum weight 
enumerators. We will show how weight enumerators also provide a means to quantify the performance of a code. For the Shor-Laflamme enumerators we have the following alternative 
to Theorem \ref{thmRains} \cite{shor}.

\begin{thm}
Let $\mathcal{Q}$ be an $((n,K))_D$ quantum code with associated 
projector $P$. Then $\mathcal{Q}$ has minimum distance of at least $d$ iff $B_i(P)=A_i(P)$
for all $0<i<d$.
\end{thm}

\noindent We conclude this section by remarking that an $((n,K,d))_D$ code is pure if and only if $B_i(P)=A_i(P)=0$ for all $0<i<d$ \cite{rains4}.

\section{Error detection}
\label{sec3}

Consider an $((n,K))_D$ quantum code $\mathcal{Q}$ with $K>1$ and associated projector $P$. Given two orthonormal 
encoded states $\ket{\phi},\ket{\psi}\in\mathcal{Q}$ ($\inner{\phi}{\psi}=0$) we know that whenever an error $E$ 
is detectable
\begin{equation}
\bra{\psi}E\ket{\psi}=\bra{\phi}E\ket{\phi}
=\left(\frac{\bra{\phi}+\bra{\psi}}{\sqrt{2}}\right)E\left(\frac{\ket{\phi}+\ket{\psi}}{\sqrt{2}}\right)=
\left(\frac{\bra{\phi}-i\bra{\psi}}{\sqrt{2}}\right)E\left(\frac{\ket{\phi}+i\ket{\psi}}{\sqrt{2}}\right)=C(E)\label{miscEQ1}
\end{equation}
where $C(E)$ is a constant depending only on $E$. From Eq. (\ref{miscEQ1}) one may easily deduce that 
$\bra{\phi}E\ket{\psi}=0$, and thus
\begin{equation}
PE\ket{\psi}=\left(P-\ket{\psi}\bra{\psi}+\ket{\psi}\bra{\psi}\right)E\ket{\psi}=C(E)\ket{\psi}
\end{equation}
since the projector $P-\ket{\psi}\bra{\psi}$ is orthogonal to $\ket{\psi}$. Consequently, 
the projective measurement $\{P,1-P\}$, which tests whether or not our corrupted state $E\ket{\psi}$ is 
in the code, will either reveal an error or project $E\ket{\psi}$ back onto the original uncorrupted state 
$\ket{\psi}$. Remarkably, in the latter case the state is in fact corrected through measurement. In a scheme 
where a corrupted state is simply discarded once an error is detected, we can estimate the rate of transmission 
(i.e. the probability that a state is accepted) with the following theorem. We first, however, define what we 
mean by a {\it random depolarizing error}.

\begin{dfn}\label{depolar}
A {\it random depolarizing error} $\Upsilon$ is a linear operator chosen 
randomly from a set $\{\Upsilon\}\subset\text{\rm End}(\mathbb{C}^N)$ with probability measure $d\mu(\Upsilon)$, and the property that
$\Exp_{\Upsilon}\left[\Upsilon^\dag A \Upsilon \right]=\tr(A)I/N$ for all linear operators 
$A\in\text{\rm End}(\mathbb{C}^N)$.
\end{dfn}

Note that if a set of random depolarizing errors $\{\Upsilon\}$ are to be considered as probabilistic quantum 
mechanical operations on a state $\rho$, then we must have the state 
$\Upsilon\rho\Upsilon^\dag/\tr\left(\Upsilon^\dag\Upsilon\rho\right)$ occurring with probability 
$\tr\left(\Upsilon^\dag\Upsilon\rho\right) d\mu(\Upsilon)$. Also note that by our definition, 
Lemma \ref{lemUps} immediately applies to random depolarizing errors. We will call $\Upsilon_S$ a random depolarizing error 
acting on qudits $S$ when $\Upsilon_S$ acts nontrivially only on qudits with indices in 
$S\subseteq\{1,\dots,n\}$ of the multi-qudit state $\ket{\psi}\in(\mathbb{C}^D)^{\otimes n}$. 
In this case $\Exp_{\Upsilon_{\! S}}\left[{\Upsilon_S}^\dag A \Upsilon_S \right]=D^{-|S|}I_S\otimes\tr_S(A)$ 
for all linear operators $A\in\text{\rm End}\left((\mathbb{C}^D)^{\otimes n}\right)$, where $I_S$ is the identity on qudits 
$S$. Some simple examples are listed below.

\begin{exm}
Let $\Upsilon_S$ be randomly chosen (with uniform probability $\mu(\Upsilon_S)=D^{-2|S|}$) from the set 
$\{\mathcal{D}(\bm{\mu},\bm{\nu})|0\leq\mu_k,\nu_k\leq D-1,\supp \mathcal{D}(\bm{\mu},\bm{\nu})\subseteq S\}$ of all 
displacement errors with support on a subset of $S$. Then $\Upsilon_S$ is a random depolarizing error acting on 
qudits $S$. This was shown in the previous section [Eq.~(\ref{displacerand})].
\end{exm}

\begin{exm} Let $\Upsilon_S$ be a unitary operator $U\in\text{U}\left(D^{|S|}\right)$ 
chosen randomly according to the Haar measure and with $\supp U=S$. From Schur's lemma we have 
$\Exp_{\supp U\subseteq S}\left[U^\dag A U\right]=D^{-|S|}I_S\otimes\tr_S(A)$ and thus 
$\Upsilon_S$ is a random depolarizing error acting on qudits $S$. 
\end{exm}

\begin{exm}\label{localexm} 
Let $\Upsilon_S$ be a local operator $\Upsilon_S=\bigotimes_{i\in S} \Upsilon_i$ with $\supp\Upsilon_S=S$, where each  
$\Upsilon_i$ is a random depolarizing error acting on qudit $i\in S$. Then  
\begin{equation}
{\textstyle\Exp_{\Upsilon_{\! S}}}\left[{\Upsilon_S}^\dag\otimes\Upsilon_S\right]=\prod_{i\in S}{\textstyle\Exp_{\Upsilon_{\! i}}}\left[{\Upsilon_i}^\dag\otimes\Upsilon_i\right]=\prod_{i\in S}\left(T_i/D\right)=D^{-|S|}T_S
\end{equation}
and by Lemma \ref{lemUps}, $\Upsilon_S$ is a random depolarizing error acting on qudits $S$. 
\end{exm}

The depolarizing channel on a single qudit $\rho$ is defined by the operation
\begin{equation}
\rho \rightarrow pI/D+(1-p)\rho 
\end{equation}
where $0\leq p\leq 1$ is the probability that the channel depolarizes the qudit. Note that, given  
a multi-qudit state $\rho$, we have
\begin{equation}
{\textstyle\Exp_{\Upsilon_{\! i}}}\left[{\Upsilon_i}^\dag \rho \Upsilon_i\right] = I_i/D\otimes\tr_i(\rho)
\end{equation}
for arbitrary random depolarizing errors $\Upsilon_i$ acting on qudit $i$. Consequently, an 
error model of $n$ depolarizing channels corrupting each individual qudit independently 
is equivalent to an error process on $\rho$ where each local 
depolarizing error $\Upsilon_S\equiv\bigotimes_{i\in S} \Upsilon_i$ ($S\subseteq\{1,\dots,n\}$)  occurs with probability 
$p^{|S|}\left(1-p\right)^{n-|S|}\prod_{i\in S}d\mu(\Upsilon_i)$. 

We now define the {\it transmission rate} to be the probability that no error is detected when an encoded state is 
corrupted by a random depolarizing error $\Upsilon$. Three different scenarios of {\it a priori} knowledge will be considered. We 
assume that either (i) the location of the corrupted qudits, $S=\supp\Upsilon$, is known, (ii) the number of 
corrupted qudits, $m=\wt\Upsilon$, is known, or (iii) the error is localized, $\Upsilon=\bigotimes_{i\in S} \Upsilon_i$, occurring 
with probability $p^{|S|}\left(1-p\right)^{n-|S|}\prod_{i\in S}d\mu(\Upsilon_i)$ as in the aforementioned depolarizing channel, and 
the probability that a single qudit is corrupted, $p$, is known. Note that in all three cases the error detection procedure remains 
the same: a projection onto the code space. The probability of successful error detection, however, will depend on our 
{\it a priori} information about the error.

\begin{dfn}
Let $\mathcal{Q}$ be an $((n,K))_D$ quantum code with associated projector $P$, and let $\Upsilon_S$ be a random 
depolarizing error acting on qudits $S\subseteq\{1,\dots,n\}$ of an encoded pure state $\ket{\psi}\in\mathcal{Q}$, 
where the state is chosen randomly. Then the {\it transmission rate on qudits $S$ under error detection}, $\T_S(P)$, is defined 
as the probability that no error is detected in the corrupted state $\ket{\psi'}=\Upsilon_S\ket{\psi}/\sqrt{\bra{\psi}{\Upsilon_S}^\dag\Upsilon_S\ket{\psi}}$, 
which occurs with probability $\bra{\psi}{\Upsilon_S}^\dag\Upsilon_S\ket{\psi}d\mu(\Upsilon_S)$. That is
\begin{equation}
\T_S(P) \equiv {\textstyle\Exp_{\Upsilon_{\! S}}}\left[\Exp_{\psi\in\mathcal{Q}}\left[\bra{\psi'}P\ket{\psi'}\bra{\psi}{\Upsilon_S}^\dag\Upsilon_S\ket{\psi}\right]\right]={\textstyle\Exp_{\Upsilon_{\! S}}}\left[\Exp_{\psi\in\mathcal{Q}}\left[\bra{\psi}{\Upsilon_S}^\dag P\Upsilon_S\ket{\psi}\right]\right].
\end{equation}
If instead the errors act on $0\leq m\leq n$ unknown qudits, the {\it transmission rate on $m$ qudits under error detection} is 
\begin{equation}
\T_m(P) \equiv \frac{m!(n-m)!}{n!}\sum_{|S|=m} \T_S(P)
\end{equation}
which is again the probability that no error is detected. Finally, the 
{\it transmission rate for the depolarizing channel under error detection} is
\begin{equation}
\T_p(P) \equiv \sum_{m=0}^n p^m(1-p)^{n-m} \frac{n!}{m!(n-m)!} \T_m(P).
\end{equation}
\end{dfn}

Now, by Lemma \ref{lemsym} and the definition of a random depolarizing error, we have
\begin{eqnarray}
{\textstyle\Exp_{\Upsilon_{\! S}}}\left[\Exp_{\psi\in\mathcal{Q}}\left[\bra{\psi}{\Upsilon_S}^\dag P\Upsilon_S\ket{\psi}\right]\right]
&=& \frac{1}{K}{\textstyle\Exp_{\Upsilon_{\! S}}}\left[\tr\left({\Upsilon_S}^\dag P \Upsilon_S P\right)\right] \\
&=& \frac{1}{D^{|S|}K}\tr\left[\left(\tr_SP\right)P\right] \\
&=& \frac{1}{D^{|S|}K}\tr_{S'}\big[(\tr_SP)^2\big]\; ,
\end{eqnarray}
and the following straightforward theorem.

\begin{thm}
Let $\mathcal{Q}$ be an $((n,K))_D$ quantum code with associated projector $P$. Then
\begin{eqnarray}
\T_S(P) &=& D^{-|S|}B_S'(P), \label{thmT1}\\
\T_m(P) &=& \frac{m!(n-m)!}{n!D^m}B_m'(P), \label{thmT2}
\end{eqnarray}
and  
\begin{equation}
\T_p(P) = \sum_{m=0}^n p^m(1-p)^{n-m} D^{-m}B_m'(P).\label{thmT3}
\end{equation}
\label{thmT}\end{thm}

Note that, given $D^{-|T|}B_T'(P)\geq D^{-|S|}B_S'(P)$ whenever $T\subseteq S$ [Eq. (\ref{rankB})], we must have  
\begin{equation}
\T_\emptyset=1\geq \T_T(P) \geq \T_S(P) \geq \frac{K}{D^n}= \T_{\{1,\dots,n\}}
\end{equation}
whenever $T\subseteq S$, and similarly
\begin{equation}
\T_0=1\geq \T_l(P)\geq \T_m(P)\geq\frac{K}{D^n}=\T_n
\end{equation}
whenever $l\leq m$. Further bounds follow from Eq.'s (\ref{boundBS}) and (\ref{boundBm}). 

The appropriate measure of success for quantum error detection is the average fidelity of all transmitted states.
Again, we consider the three different cases of {\it a priori} information about the error.

\begin{dfn}
Let $\mathcal{Q}$ be an $((n,K))_D$ quantum code with associated projector $P$, and let $\Upsilon_S$ be a random 
depolarizing error acting on qudits $S\subseteq\{1,\dots,n\}$ of an encoded pure state $\ket{\psi}\in\mathcal{Q}$, where 
the state is chosen randomly. Then the {\it transmission fidelity on qudits $S$ under error detection}, $\Fd_S(P)$, is 
defined as the average fidelity of all transmitted states, given corrupted input states of the form 
$\ket{\psi'}=\Upsilon_S\ket{\psi}/\sqrt{\bra{\psi}{\Upsilon_S}^\dag\Upsilon_S\ket{\psi}}$, which occur with probability 
$\bra{\psi}{\Upsilon_S}^\dag\Upsilon_S\ket{\psi}d\mu(\Upsilon_S)$. That is
\begin{equation}
\Fd_S(P) \equiv \frac{{\textstyle\Exp_{\Upsilon_{\! S}}}\left[\Exp_{\psi\in\mathcal{Q}}\left[\inner{\phi}{\psi}\inner{\psi}{\phi}\bra{\psi'}P\ket{\psi'}\bra{\psi}{\Upsilon_S}^\dag\Upsilon_S\ket{\psi}\right]\right]}{\T_S(P)}=\frac{{\textstyle\Exp_{\Upsilon_{\! S}}}\left[\Exp_{\psi\in\mathcal{Q}}\left[\bra{\psi}{\Upsilon_S}^\dag\ket{\psi}\bra{\psi}\Upsilon_S\ket{\psi}\right]\right]}{\T_S(P)}
\end{equation}
where $\ket{\phi}=P\ket{\psi'}/\sqrt{\bra{\psi'}P\ket{\psi'}}$. If instead the errors act on $0\leq m\leq n$ unknown qudits, the 
{\it transmission fidelity on $m$ qudits under error detection} is 
\begin{equation}
\Fd_m(P) \equiv \frac{m!(n-m)!}{n!\T_m(P)}\sum_{|S|=m} \T_S(P)\Fd_S(P)
\end{equation}
which is again the average fidelity of all transmitted states. Finally, the 
{\it transmission fidelity for the depolarizing channel under error detection} is
\begin{equation}
\Fd_p(P) \equiv \frac{1}{\T_p(P)}\sum_{m=0}^n p^m(1-p)^{n-m} \frac{n!}{m!(n-m)!} \T_m(P)\Fd_m(P).
\end{equation}
\label{detectfideldfn}
\end{dfn}

By Lemmas \ref{lemsym} and \ref{lemUps}, and the definition of a random depolarizing error, we have
\begin{eqnarray}
{\textstyle\Exp_{\Upsilon_{\! S}}}\left[\Exp_{\psi\in\mathcal{Q}}\left[\bra{\psi}{\Upsilon_S}^\dag\ket{\psi}\bra{\psi}\Upsilon_S\ket{\psi}\right]\right]
&=& \tr\left({\textstyle\Exp_{\Upsilon_{\! S}}}\left[{\Upsilon_S}^\dag\otimes\Upsilon_S\right]\Exp_{\psi\in\mathcal{Q}}\left[\ket{\psi}\bra{\psi}\otimes\ket{\psi}\bra{\psi}\right]\right) \\
&=& \frac{1}{D^{|S|}K(K+1)}\tr\left[T_S\left(P\otimes P\right)\left(1+T\right)\right] \\
&=& \frac{1}{D^{|S|}K(K+1)}\tr\left[\left(P\otimes P\right)(T_S+T_{S'})\right] \\
&=& \frac{1}{D^{|S|}K(K+1)}\left(\tr_S\left[(\tr_{S'}P)^2\right]+\tr_{S'}\left[(\tr_SP)^2\right]\right)
\end{eqnarray}
and another straightforward result: 

\begin{thm}
Let $\mathcal{Q}$ be an $((n,K))_D$ quantum code with associated projector $P$. Then
\begin{eqnarray}
\Fd_S(P) &=& \frac{KA_S'(P)+B_S'(P)}{(K+1)B_S'(P)}, \label{thmF1} \\
\Fd_m(P) &=& \frac{KA_m'(P)+B_m'(P)}{(K+1)B_m'(P)}, \label{thmF2}
\end{eqnarray}
and
\begin{equation}
\Fd_p(P)= \frac{1}{(K+1)\T_p(P)}\sum_{m=0}^n p^m(1-p)^{n-m} D^{-m}\left[KA_m'(P)+B_m'(P)\right]. \label{thmF3}
\end{equation}
\label{thmF}\end{thm}

Given $KA_S'(P)=B_{S'}'(P)\geq A_{S'}'(P)=B_S'(P)/K$ we find that  
\begin{equation}
\Fd_\emptyset=1\geq \Fd_S(P) \geq \frac{1}{K}= \Fd_{\{1,\dots,n\}}
\end{equation}
and similarly
\begin{equation}
\Fd_0=1\geq \Fd_m(P)\geq\frac{1}{K}=\Fd_n.
\end{equation}
Unlike the transmission rate, the fidelity need not be monotonic. The relation $B_m'(P)=KA_{n-m}'(P)$ implies that 
$(K+1)\Fd_{n-m}(P)-1=1/\left[(K+1)\Fd_m(P)-1\right]$, and
in particular, $\Fd_{n/2}(P)=2/(K+1)$ whenever $n$ is even.

We can interpret the fidelity as the probability of measuring the original encoded state
$\ket{\psi}$ after a projective measurement $\{\ket{\psi}\bra{\psi},1-\ket{\psi}\bra{\psi}\}$ 
of the transmitted state. The failure rate for the code under error detection will be defined 
as the probability that an error is not detected and this final measurement reveals a negative 
outcome. That is, the {\it failure rate on qudits $S$ under error detection}, 
\begin{equation}
\fd_S(P)\equiv\T_S(P)\left[1-\Fd_S(P)\right]=\frac{K}{D^{|S|}(K+1)}\left[B_S'(P)-A_S'(P)\right],
\end{equation}
the {\it failure rate on $m$ qudits under error detection} 
\begin{equation}
\fd_m(P)\equiv\T_m(P)\left[1-\Fd_m(P)\right]=\frac{m!(n-m)!K}{n!D^m(K+1)}\left[B_m'(P)-A_m'(P)\right],
\end{equation}
and the {\it failure rate for the depolarizing channel under error detection}
\begin{equation}
\fd_p(P) \equiv \T_p(P)\left[1-\Fd_p(P)\right]=\frac{K}{K+1}\sum_{m=0}^n p^m(1-p)^{n-m} D^{-m}\left[B_m'(P)-A_m'(P)\right].
\end{equation}
Using relations (\ref{rel1}) and (\ref{rel2}) we find that
\begin{eqnarray}
\fd_p(P) &=& \frac{K}{K+1}\sum_{m=0}^n p^m(1-p)^{n-m} D^{-m} \sum_{i=0}^m\frac{(n-i)!}{(m-i)!(n-m)!}\left[B_i(P)-A_i(P)\right] \\
&=& \frac{K}{K+1}\sum_{i=0}^n \sum_{m=i}^n p^m(1-p)^{n-m}D^{-m}\frac{(n-i)!}{(m-i)!(n-m)!} \left[B_i(P)-A_i(P)\right] \\
&=& \frac{K}{K+1}\sum_{i=0}^n \bigg(\frac{p}{D^2}\bigg)^i\left(1-\frac{D^2-1}{D^2}p\right)^{n-i}\left[B_i(P)-A_i(P)\right]
\end{eqnarray}
which under the transformation $p \rightarrow pD^2/(D^2-1)$ agrees with Ashikhmin {\it et al}~\cite{ashikhmin2}. Similarly, the quantities 
$\T_p$ and $\Fd_p$ may also be converted into sums over the Shor-Laflamme enumerators.

Finally, when $p$ is small, for an $((n,K,d))_D$ code we find that 
\begin{eqnarray}
\Fd_p(P) &=& 1-\fd_p(P)+O\left(p^{d+1}\right) \\
&=& 1 - \frac{n!\fd_d(P)}{d!(n-d)!} p^d + O\left(p^{d+1}\right) \label{detectsmallp}
\end{eqnarray}
and thus we consider the quantity
\begin{equation}
c\equiv\frac{n!\fd_d(P)}{d!(n-d)!}=\frac{K\left[B_d'(P)-A_d'(P)\right]}{(K+1)D^d}=\frac{K\left[B_d(P)-A_d(P)\right]}{(K+1)D^{2d}} \label{detectsmallpc}
\end{equation}
to be a useful second-order parameter when comparing codes of the same minimum distance.

\section{Error correction}
\label{sec4}

Error correction is achieved through a two-step process: a projective measurement followed by
a unitary operation conditioned on the measurement outcome. In Section \ref{sec3} error detection was treated 
for arbitrary quantum codes. We will restrict our analysis, however, to stabilizer codes with $D$ a prime power for the 
case of error correction. 

When $D=D'$, prime, a {\it stabilizer code} \cite{calderbank,gottesman,rains2,ashikhmin,grassl2} is defined as a joint eigenspace 
of an Abelian subgroup $\mathcal{S}$ (called the {\it stabilizer}) of the {\it error group} 
\begin{equation}
\mathcal{E}\equiv\left\{e^{2\pi i\gamma/D}e^{i\pi\bm{\mu}\cdot\bm{\nu}}\mathcal{D}(\bm{\mu},\bm{\nu})\,\Big|\,0\leq\mu_j,\nu_j,\gamma\leq D-1\,,\, 1\leq j\leq n\right\}. 
\end{equation}
We also assume that the center of $\mathcal{E}$ is contained in the stabilizer i.e. 
$\mathcal{S}\supseteq\mathcal{Z}\equiv\left\{e^{2\pi i\gamma/D}I\,|\,0\leq\gamma\leq D-1\right\}$. When this is not the case we may 
simply extend $\mathcal{S}$ by $\mathcal{Z}$. The dimension of the code space is then $K=D^k$, where $k$ is an integer such that 
$|\mathcal{S}|=D^{n-k+1}$. The code has a minimum distance of at least $d$ if there are no elements of weight $<d$ in 
$C(\mathcal{S})\backslash\mathcal{S}$ where the centralizer 
$C(\mathcal{S})\equiv\left\{E\in\mathcal{E}\,|\,EF=FE \;\,\forall F\in\mathcal{S}\right\}$. We use the notation $[[n,k,d]]_D$ for stabilizer codes, 
or simply $[[n,k,d]]$ when $D=2$. 

When $D=D'^l$, a prime power, we instead use the error group 
\begin{equation}
\mathcal{E}\equiv\left\{e^{2\pi i\gamma/D'}e^{i\pi\bm{\mu}\cdot\bm{\nu}}\mathcal{D}(\bm{\mu},\bm{\nu})\,\Big|\,0\leq\mu_j,\nu_j,\gamma\leq D'-1\,,\, 1\leq j\leq ln\right\} 
\end{equation}
where the first $l$ displacement operators in the tensor product $\mathcal{D}(\bm{\mu},\bm{\nu})$ act on the first qudit, and so on.
Our definition of a stabilizer code \cite{ashikhmin} is then unchanged except that the dimension of the code space, 
$K=D'^{ln-r}\equiv D^k$, where $r$ is an integer such that $|\mathcal{S}|=D'^{r+1}$. Thus $k=n-r/l$ may now be non-integer. 

The process of error correction for stabilizer codes is initiated by a measurement with orthogonal projectors in the form
\begin{equation}
P_{\lambda}\equiv \frac{1}{|\mathcal{S}|}\sum_{E\in\mathcal{S}}\lambda(E)^{-1}E \qquad\qquad \left(\;\sum_{\lambda}P_{\lambda}=I \;\right) \label{Plambda}
\end{equation}
where $\lambda(E)$ is an eigenvalue associated with $E$ i.e. $P_{\lambda}EP_{\lambda}=\lambda(E)P_{\lambda}$. Since $D'$ is prime each member of 
$\mathcal{S}\backslash\mathcal{Z}$ has $D'$ distinct eigenvalues in the form $e^{2\pi i\gamma/D'}$ ($0\leq\gamma\leq D'-1$) and thus there 
are $D'^r=D^{n-k}$ distinct functions $\lambda$, each denoting an orthogonal subspace of dimension $D^k$. More precisely, 
$\lambda:\mathcal{S}\rightarrow\mathbb{C}$ is one of the $D'^r$ distinct {\it characters} of the Abelian group 
$\mathcal{S}$ satisfying $\lambda(e^{2\pi i/D'}I)=e^{2\pi i/D'}$ \cite{ashikhmin}.
One such character, $\lambda_0$ say, labels the projector $P$ of the code space itself. A measurement result of $\lambda$ will project 
the encoded state into the subspace defined by $P_\lambda$. An error is detected when $\lambda\neq\lambda_0$ and we attempt 
correction. 

From Eq. (\ref{Plambda}) and (\ref{AS2})-(\ref{BS3}) we find that for stabilizer codes
\begin{eqnarray}
A_S'(P) &=& \frac{1}{D^{|S|}D'}\left|\left\{E\in\mathcal{S}\,\big|\,\supp E\subseteq S\right\}\right| \\
B_S'(P) &=& \frac{1}{D^{|S|}D'}\left|\left\{E\in C(\mathcal{S})\,\big|\,\supp E\subseteq S\right\}\right|.
\end{eqnarray}
Similarly
\begin{eqnarray}
A_m(P) &=& \frac{1}{D'}\left|\left\{E\in\mathcal{S}\,\big|\,\wt E=m\right\}\right| \\
B_m(P) &=& \frac{1}{D'}\left|\left\{E\in C(\mathcal{S})\,\big|\,\wt E=m\right\}\right|
\end{eqnarray}
and the weight enumerators $A_m'(P)$ and $B_m'(P)$ may be found through Eq.'s (\ref{Amp}) and (\ref{Bmp}) or Eq.'s (\ref{rel1}) and (\ref{rel2}).

Define the subsets 
\begin{eqnarray}
\mathcal{E}_{\lambda} &\equiv& \left\{E\in\mathcal{E}\,\Big|\,\lambda_0(F)E^\dag F E=\lambda(F) F \;\,\forall\, F\in\mathcal{S}\right\} \\
&=& \left\{E\in\mathcal{E}\,\Big|\,E^\dag P_{\lambda} E=P\right\} \\
\mathcal{E}_{\lambda,S} &\equiv& \left\{E\in\mathcal{E}_{\lambda}\,\Big|\, \supp E\subseteq S\right\}.
\end{eqnarray}
The sets $\mathcal{E}_{\lambda}$ are disjoint with $\mathcal{E}=\bigcup_{\lambda}\mathcal{E}_{\lambda}$. The measurement part 
of the error correction process is then described through the following lemma.

\begin{lem}
Let $\mathcal{Q}$ be an $[[n,k]]_D$ quantum code with associated projector $P$, and let $\Upsilon_S$ be a random 
depolarizing error acting on qudits $S\subseteq\{1,\dots,n\}$ of an encoded pure state $\ket{\psi}\in\mathcal{Q}$, 
where the state is chosen randomly. The probability of result $\lambda$ under a projective 
measurement with elements (\ref{Plambda}) is
\begin{eqnarray}
\prob(\lambda) &\equiv& {\textstyle\Exp_{\Upsilon_{\! S}}}\left[\Exp_{\psi\in\mathcal{Q}}\left[\bra{\psi}{\Upsilon_S}^\dag P_{\lambda}\Upsilon_S\ket{\psi}\right]\right] \\
&=& \frac{1}{D^{|S|}K}\tr_{S'}\big[(\tr_SP)(\tr_SP_{\lambda})\big] \label{prob2}\\
&=& \left\{\begin{array}{ll} \T_S(P) & \text{\rm if }\quad \mathcal{E}_{\lambda,S} \neq \emptyset \\ 0 & \text{\rm if }\quad \mathcal{E}_{\lambda,S} = \emptyset \end{array}\right. . \label{prob3}
\end{eqnarray}
\label{lem15}\end{lem}
\begin{prf}
Eq. (\ref{prob2}) may be shown through a simple variation of the proof of Theorem \ref{thmT}. Now, if $\mathcal{E}_{\lambda,S}=\emptyset$ 
then ${\Upsilon_S}^\dag P_{\lambda}\Upsilon_S$ is necessarily orthogonal to all $\ket{\psi}\in\mathcal{Q}$ since $\supp\Upsilon_S\subseteq S$, 
and hence $\prob(\lambda)=0$. Otherwise, if 
$\mathcal{E}_{\lambda,S}\neq\emptyset$ we find that 
$\tr_{S'}\big[(\tr_SP)(\tr_SP_{\lambda})\big]=\tr_{S'}\big[(\tr_SP)(\tr_S EPE^\dag)\big]=\tr_{S'}\big[(\tr_SP)^2\big]$ where 
$E\in\mathcal{E}_{\lambda,S}$, and Eq. (\ref{prob3}) follows.
\end{prf}

For stabilizer codes, when an error is detected with result $\lambda$, we attempt correction by applying some unitary 
${C_{\lambda}}^\dag$ where $C_{\lambda}\in\mathcal{E}_{\lambda}$. Our task is to find the optimal $C_{\lambda}$ 
for which the output fidelity is maximized. This operator will, in general, depend on our {\it a priori} information about the error.
As in the case of error detection, we consider three possibilities: (i) the location of the corrupted qudits, $S=\supp\Upsilon$, is known, (ii) the number of 
corrupted qudits, $m=\wt\Upsilon$, is known, or (iii) the error is localized, $\Upsilon=\bigotimes_{i\in S} \Upsilon_i$, and occurs 
with probability $p^{|S|}\left(1-p\right)^{n-|S|}\prod_{i\in S}d\mu(\Upsilon_i)$, where the probability that a single qudit is corrupted, $p$, is known.

\begin{dfn}
Let $\mathcal{Q}$ be an $[[n,k]]_D$ quantum code with associated projector $P$, and let $\Upsilon_S$ be a random 
depolarizing error acting on qudits $S\subseteq\{1,\dots,n\}$ of an encoded pure state $\ket{\psi}\in\mathcal{Q}$, where 
the state is chosen randomly. Then the {\it transmission fidelity on qudits $S$ under error correction}, 
$\Fc_S(P)$, is defined as the maximum possible average fidelity of all transmitted states under error correction, given input 
states of the form $\ket{\psi'}=\Upsilon_S\ket{\psi}/\sqrt{\bra{\psi}{\Upsilon_S}^\dag\Upsilon_S\ket{\psi}}$, which occur with probability 
$\bra{\psi}{\Upsilon_S}^\dag\Upsilon_S\ket{\psi}d\mu(\Upsilon_S)$. That is
\begin{eqnarray}
\Fc_S(P) &\equiv& \sum_{\lambda}\max_{C_{\lambda}\in\mathcal{E}_{\lambda}}{\textstyle\Exp_{\Upsilon_{\! S}}}\left[\Exp_{\psi\in\mathcal{Q}}\left[\inner{\phi}{\psi}\inner{\psi}{\phi}\bra{\psi'}P_{\lambda}\ket{\psi'}\bra{\psi}{\Upsilon_S}^\dag\Upsilon_S\ket{\psi}\right]\right] \\
&=& \sum_{\lambda}\max_{C_{\lambda}\in\mathcal{E}_{\lambda}}{\textstyle\Exp_{\Upsilon_{\! S}}}\left[\Exp_{\psi\in\mathcal{Q}}\left[\bra{\psi}{C_{\lambda}}^\dag\Upsilon_S\ket{\psi}\bra{\psi}{\Upsilon_S}^\dag C_{\lambda}\ket{\psi}\right]\right] \label{misceq5}
\end{eqnarray}
where $\ket{\phi}={C_{\lambda}}^\dag P_{\lambda}\ket{\psi'}/\sqrt{\bra{\psi'}P_{\lambda}\ket{\psi'}}$. If instead the errors act on 
$0\leq m\leq n$ unknown qudits, the {\it transmission fidelity on $m$ qudits under error correction} is
\begin{equation}
\Fc_m(P) \equiv \frac{m!(n-m)!}{n!}\sum_{\lambda}\max_{C_{\lambda}\in\mathcal{E}_{\lambda}}\sum_{|S|=m}{\textstyle\Exp_{\Upsilon_{\! S}}}\left[\Exp_{\psi\in\mathcal{Q}}\left[\bra{\psi}{C_{\lambda}}^\dag\Upsilon_S\ket{\psi}\bra{\psi}{\Upsilon_S}^\dag C_{\lambda}\ket{\psi}\right]\right] 
\end{equation}
which is again the maximum possible average fidelity of all transmitted states. Finally, the 
{\it transmission fidelity for the depolarizing channel under error correction} is
\begin{equation}
\Fc_p(P) \equiv \sum_{\lambda}\max_{C_{\lambda}\in\mathcal{E}_{\lambda}}\sum_{m=0}^n p^m(1-p)^{n-m}\sum_{|S|=m}{\textstyle\Exp_{\Upsilon_{\! S}}}\left[\Exp_{\psi\in\mathcal{Q}}\left[\bra{\psi}{C_{\lambda}}^\dag\Upsilon_S\ket{\psi}\bra{\psi}{\Upsilon_S}^\dag C_{\lambda}\ket{\psi}\right]\right] .
\end{equation}
\end{dfn}

\begin{thm}
Let $\mathcal{Q}$ be an $[[n,k]]_D$ quantum code with associated projector $P$. Then
\begin{equation}
\Fc_S(P)=\Fd_S(P)=\frac{KA_S'(P)+B_S'(P)}{(K+1)B_S'(P)}.
\end{equation}
\label{thm17}\end{thm}

\begin{prf}
First note that
\begin{equation}
\bra{\psi}U^\dag\big(I_S\otimes\tr_S\ket{\psi}\bra{\psi}\big)U\ket{\psi}\leq\bra{\psi}\big(I_S\otimes\tr_S\ket{\psi}\bra{\psi}\big)\ket{\psi}
\label{lem17}\end{equation} 
for all product unitaries $U=U_S\otimes U_{S'}$ where $\supp U_S\subseteq S$ and $\supp U_{S'}\subseteq S'$. We can show this by 
rewriting (\ref{lem17}) as
\begin{equation}
\tr_{S'}\left[U_{S'}^\dag\big(\tr_S\ket{\psi}\bra{\psi}\big)U_{S'}\big(\tr_S\ket{\psi}\bra{\psi}\big)\right]\leq\tr_{S'}\left[\big(\tr_S\ket{\psi}\bra{\psi}\big)^2\right].
\end{equation} 
Now with $A=U_{S'}^\dag\big(\tr_S\ket{\psi}\bra{\psi}\big)U_{S'}$ and $B=\tr_S\ket{\psi}\bra{\psi}$ we see that (\ref{lem17}) follows 
from the Cauchy-Schwarz inequality: $|\tr AB^\dag|^2\leq\tr AA^\dag\tr BB^\dag$.

By the definition of a random depolarizing error 
\begin{eqnarray}
{\textstyle\Exp_{\Upsilon_{\! S}}}\left[\Exp_{\psi\in\mathcal{Q}}\left[\bra{\psi}{C_{\lambda}}^\dag\Upsilon_S\ket{\psi}\bra{\psi}{\Upsilon_S}^\dag C_{\lambda}\ket{\psi}\right]\right] &=& D^{-|S|}\Exp_{\psi\in\mathcal{Q}}\left[\bra{\psi}{C_{\lambda}}^\dag \big(I_S\otimes\tr_S\ket{\psi}\bra{\psi}\big) C_{\lambda}\ket{\psi}\right] \label{eqA}\\
& \leq & D^{-|S|}\Exp_{\psi\in\mathcal{Q}}\left[\bra{\psi} \big(I_S\otimes\tr_S\ket{\psi}\bra{\psi}\big) \ket{\psi}\right] \label{eq17}\\
&=& {\textstyle\Exp_{\Upsilon_{\! S}}}\left[\Exp_{\psi\in\mathcal{Q}}\left[\bra{\psi}\Upsilon_S\ket{\psi}\bra{\psi}{\Upsilon_S}^\dag \ket{\psi}\right]\right] \\
&=& \T_S(P)\Fd_S(P) \\
&=& \frac{KA_S'(P)+B_S'(P)}{(K+1)D^{|S|}}
\end{eqnarray}
where we have used Eq. (\ref{lem17}) for the inequality (\ref{eq17}) and Theorems \ref{thmF} and \ref{thmT}. Now since 
$\supp\Upsilon_S\subseteq S$, with $\psi\in\mathcal{Q}$ and $C_{\lambda}\in\mathcal{E}_{\lambda}$, the two states
$\Upsilon_S\ket{\psi}$ and $C_{\lambda}\ket{\psi}$ are necessarily orthogonal whenever $\mathcal{E}_{\lambda,S}=\emptyset$.
Consequently
\begin{eqnarray}
\Fc_S(P) &=& \mathop{\sum_{\lambda}}_{\mathcal{E}_{\lambda,S}\neq\emptyset}\max_{C_{\lambda}\in\mathcal{E}_{\lambda}}{\textstyle\Exp_{\Upsilon_{\! S}}}\left[\Exp_{\psi\in\mathcal{Q}}\left[\bra{\psi}{C_{\lambda}}^\dag\Upsilon_S\ket{\psi}\bra{\psi}{\Upsilon_S}^\dag C_{\lambda}\ket{\psi}\right]\right] \\
&\leq& \frac{KA_S'(P)+B_S'(P)}{(K+1)D^{|S|}}\mathop{\sum_{\lambda}}_{\mathcal{E}_{\lambda,S}\neq\emptyset} 1 \label{eq17b}\\
&=& \frac{KA_S'(P)+B_S'(P)}{(K+1)D^{|S|}}\sum_{\lambda} \frac{\tr_{S'}\big[(\tr_SP)(\tr_SP_{\lambda})\big]}{\tr_{S'}\big[(\tr_SP)^2\big]} \label{eqB}\\
&=& \frac{KA_S'(P)+B_S'(P)}{(K+1)D^{|S|}}\frac{D^{|S|}K}{\tr_{S'}\big[(\tr_SP)^2\big]} \\
&=& \frac{KA_S'(P)+B_S'(P)}{(K+1)B_S'(P)} \label{eqC}
\end{eqnarray}
where Lemma \ref{lem15} was used for Eq. (\ref{eqB}). We have thus found a bound for the fidelity $\Fc_S(P)$. 
In fact, this bound is reached by simply choosing any $C_{\lambda}\in\mathcal{E}_{\lambda,S}\subseteq\mathcal{E}_{\lambda}$.
Then $\supp C_{\lambda}\subseteq S$, and hence, the inequalities (\ref{eq17}) and (\ref{eq17b}) are saturated, giving the desired 
result.
\end{prf}

We now extend the definition of our weight enumerators to arbitrary operators:
\begin{eqnarray}
A_S'(A) &\equiv& \frac{1}{D^{|S|}K^2}\sum_{\supp \mathcal{D}(\bm{\mu},\bm{\nu})\subseteq S} \tr\left[\mathcal{D}(\bm{\mu},\bm{\nu})^\dag A\right] \tr\left[\mathcal{D}(\bm{\mu},\bm{\nu})A^\dag\right] \label{extASp}\\
&=& \frac{1}{K^2}\tr_S\left[(\tr_{S'}A^\dag)(\tr_{S'}A)\right] \\
B_S'(A) &\equiv&\frac{1}{D^{|S|}K}\sum_{\supp \mathcal{D}(\bm{\mu},\bm{\nu})\subseteq S} \tr\left[\mathcal{D}(\bm{\mu},\bm{\nu})^\dag A\mathcal{D}(\bm{\mu},\bm{\nu})A^\dag\right] \label{extBSp}\\
&=& \frac{1}{K}\tr_{S'}\left[(\tr_S A^\dag)(\tr_S A)\right].
\end{eqnarray}
The enumerators $A_m'(A)$ and $B_m'(A)$ are then extended through Eq.'s (\ref{Amp}) and (\ref{Bmp}), and enumerators $A_m(A)$ and $B_m(A)$ 
in the same manner as Eq.'s (\ref{extASp}) and (\ref{extBSp}).

\begin{prp}
Let $\mathcal{Q}$ be an $[[n,k]]_D$ quantum code with associated projector $P$. Then
\begin{equation}
\Fc_m(P)=\frac{1}{K+1}+\frac{m!(n-m)!K}{n!D^m(K+1)}\sum_{\lambda}\max_{C_{\lambda}\in\mathcal{E}_{\lambda}}A_m'\left(C_{\lambda}P\right) \label{prop1}
\end{equation}
and
\begin{equation}
\Fc_p(P)=\frac{1}{K+1}+\frac{K}{K+1}\sum_{\lambda}\max_{C_{\lambda}\in\mathcal{E}_{\lambda}}\sum_{m=0}^n p^m(1-p)^{n-m} D^{-m}A_m'\left(C_{\lambda}P\right). \label{prop2}
\end{equation}
\label{prop}\end{prp}

\begin{prf}
By Lemmas \ref{lemsym} and \ref{lemUps}, and the definition of a random depolarizing error we easily obtain
\begin{equation}
{\textstyle\Exp_{\Upsilon_{\! S}}}\left[\Exp_{\psi\in\mathcal{Q}}\left[\bra{\psi}{C_{\lambda}}^\dag\Upsilon_S\ket{\psi}\bra{\psi}{\Upsilon_S}^\dag C_{\lambda}\ket{\psi}\right]\right]=\frac{\tr_S\left[\big(\tr_{S'}P {C_{\lambda}}^\dag\big)\big(\tr_{S'}C_{\lambda}P\big)\right]+\tr_{S'}\left[\big(\tr_S C_{\lambda}P {C_{\lambda}}^\dag\big)\big(\tr_S P\big)\right]}{D^{|S|}K(K+1)} \label{misceq3}
\end{equation}
and given that $C_{\lambda}P {C_{\lambda}}^\dag=P_{\lambda}$ for all $C_{\lambda}\in\mathcal{E}_{\lambda}$, 
we have
\begin{eqnarray}
\sum_{\lambda}\max_{C_{\lambda}\in\mathcal{E}_{\lambda}}\tr_{S'}\left[\Big(\tr_S C_{\lambda}P {C_{\lambda}}^\dag\Big)\Big(\tr_S P\Big)\right] &=& \sum_{\lambda}\tr_{S'}\left[\big(\tr_S P_{\lambda}\big)\big(\tr_S P\big)\right] \\
&=&\tr_{S'}\left[\big(\tr_S I\big)\big(\tr_S P\big)\right] \\
&=& D^{|S|}K \label{prfb}
\end{eqnarray}
and hence, Eq. (\ref{prop1}). The proof of Eq. (\ref{prop2}) is similar.
\end{prf}
\begin{figure}[t]
\includegraphics[scale=1]{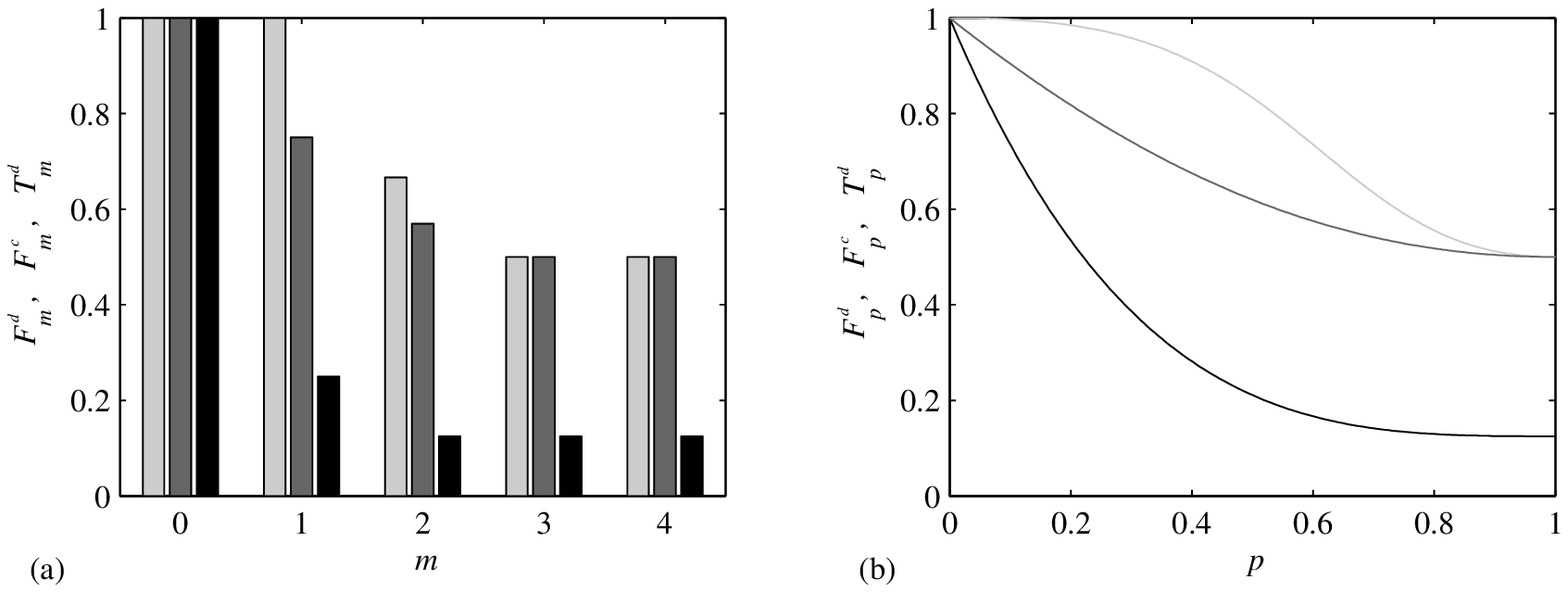}
\caption{For the $[[4,1,2]]$ code $\mathcal{G}_{4a}$ we plot (a) from left to right, the transmission fidelity on $m$ 
qubits under error detection $\Fd_m$ [Eq.~(\ref{thmF2})] (light), the transmission fidelity on $m$ qubits under error correction $\Fc_m$ [Eq.~(\ref{prop1})] (dark), the transmission rate 
on $m$ qubits under error detection $\T_m$ [Eq.~(\ref{thmT2})] (black), and (b) from top to bottom, $\Fd_p$ [Eq.~(\ref{thmF3})] (light), $\Fc_p$ [Eq.~(\ref{prop2})] (dark), and $\T_p$ [Eq.~(\ref{thmT3})] (black) for the depolarizing channel.} 
\label{fig1}
\end{figure}
\begin{figure}[t]
\includegraphics[scale=1]{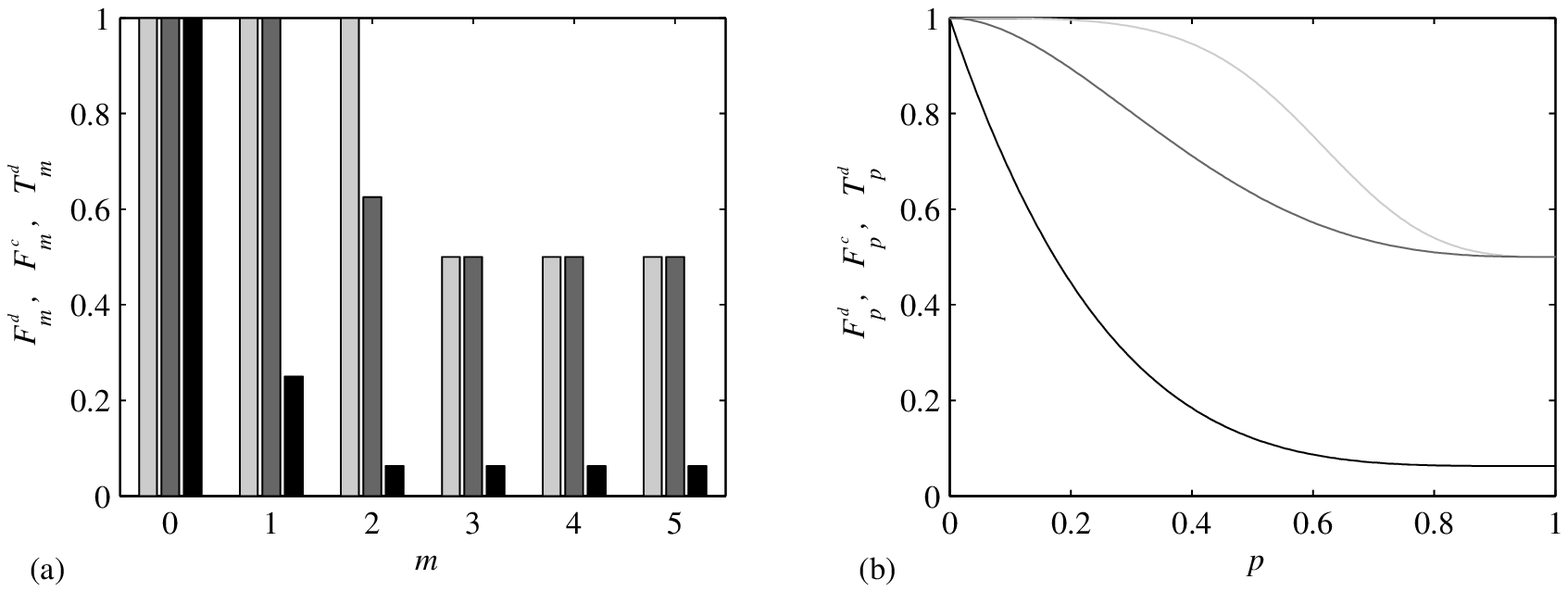}
\caption{For the $[[5,1,3]]$ code $\mathcal{G}_{5}$ we plot (a) from left to right, the transmission fidelity on $m$ 
qubits under error detection $\Fd_m$ [Eq.~(\ref{thmF2})] (light), the transmission fidelity on $m$ qubits under error correction $\Fc_m$ [Eq.~(\ref{prop1})] (dark), the transmission rate 
on $m$ qubits under error detection $\T_m$ [Eq.~(\ref{thmT2})] (black), and (b) from top to bottom, $\Fd_p$ [Eq.~(\ref{thmF3})] (light), $\Fc_p$ [Eq.~(\ref{prop2})] (dark), and $\T_p$ [Eq.~(\ref{thmT3})] (black) for the depolarizing channel.} 
\label{fig2}
\end{figure}
\begin{figure}[t]
\includegraphics[scale=1]{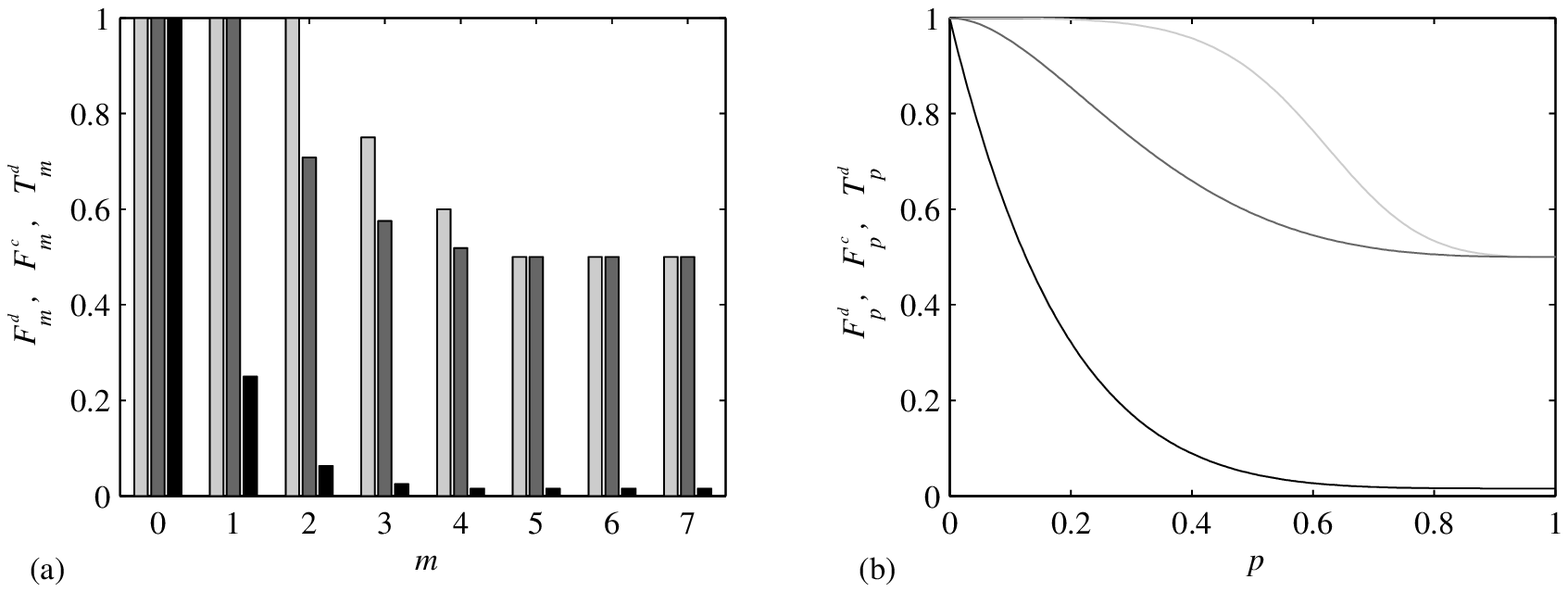}
\caption{For Steane's $[[7,1,3]]$ code $\mathcal{G}_{7b}$ we plot (a) from left to right, the transmission fidelity on $m$ 
qubits under error detection $\Fd_m$ [Eq.~(\ref{thmF2})] (light), the transmission fidelity on $m$ qubits under error correction $\Fc_m$ [Eq.~(\ref{prop1})] (dark), the transmission rate 
on $m$ qubits under error detection $\T_m$ [Eq.~(\ref{thmT2})] (black), and (b) from top to bottom, $\Fd_p$ [Eq.~(\ref{thmF3})] (light), $\Fc_p$ [Eq.~(\ref{prop2})] (dark), and $\T_p$ [Eq.~(\ref{thmT3})] (black) for the depolarizing channel.} 
\label{fig3}
\end{figure}
\begin{figure}[t]
\includegraphics[scale=1]{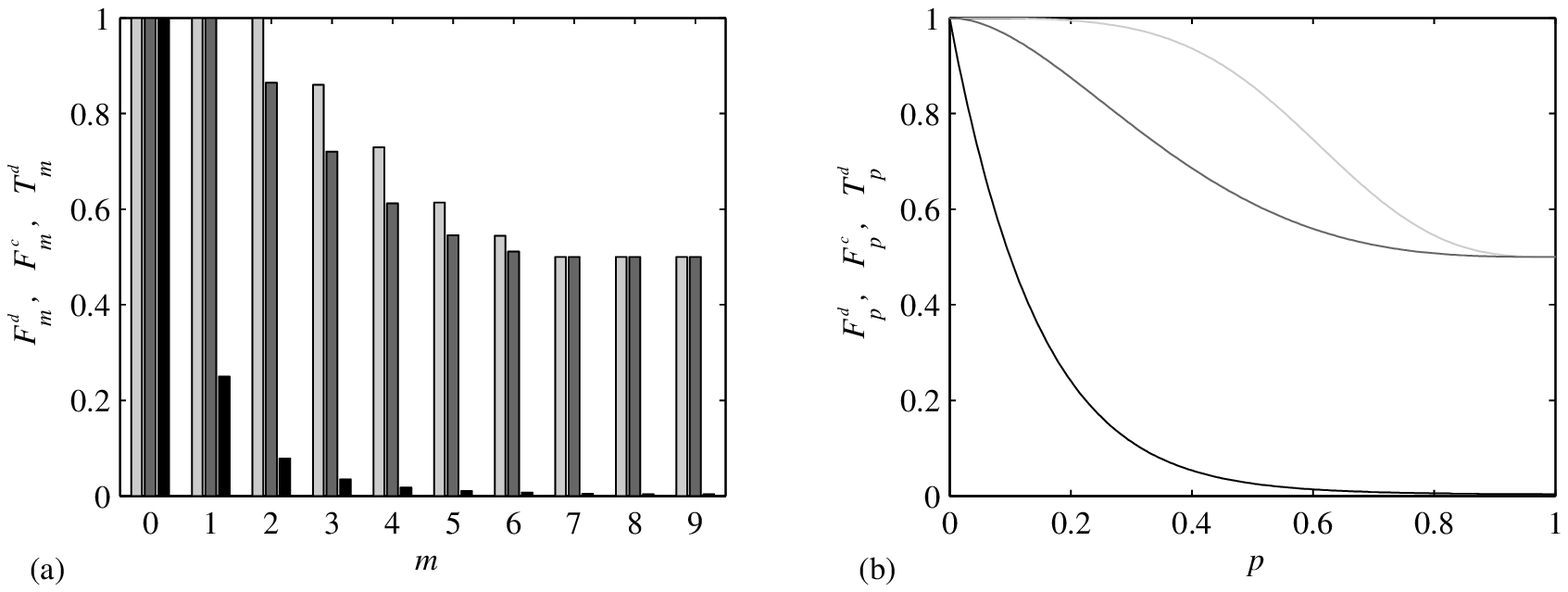}
\caption{For Shor's $[[9,1,3]]$ code $\mathcal{G}_{9c}$ we plot (a) from left to right, the transmission fidelity on $m$ 
qubits under error detection $\Fd_m$ [Eq.~(\ref{thmF2})] (light), the transmission fidelity on $m$ qubits under error correction $\Fc_m$ [Eq.~(\ref{prop1})] (dark), the transmission rate 
on $m$ qubits under error detection $\T_m$ [Eq.~(\ref{thmT2})] (black), and (b) from top to bottom, $\Fd_p$ [Eq.~(\ref{thmF3})] (light), $\Fc_p$ [Eq.~(\ref{prop2})] (dark), and $\T_p$ [Eq.~(\ref{thmT3})] (black) for the depolarizing channel.} 
\label{fig4}
\end{figure}
\begin{figure}[t]
\includegraphics[scale=1]{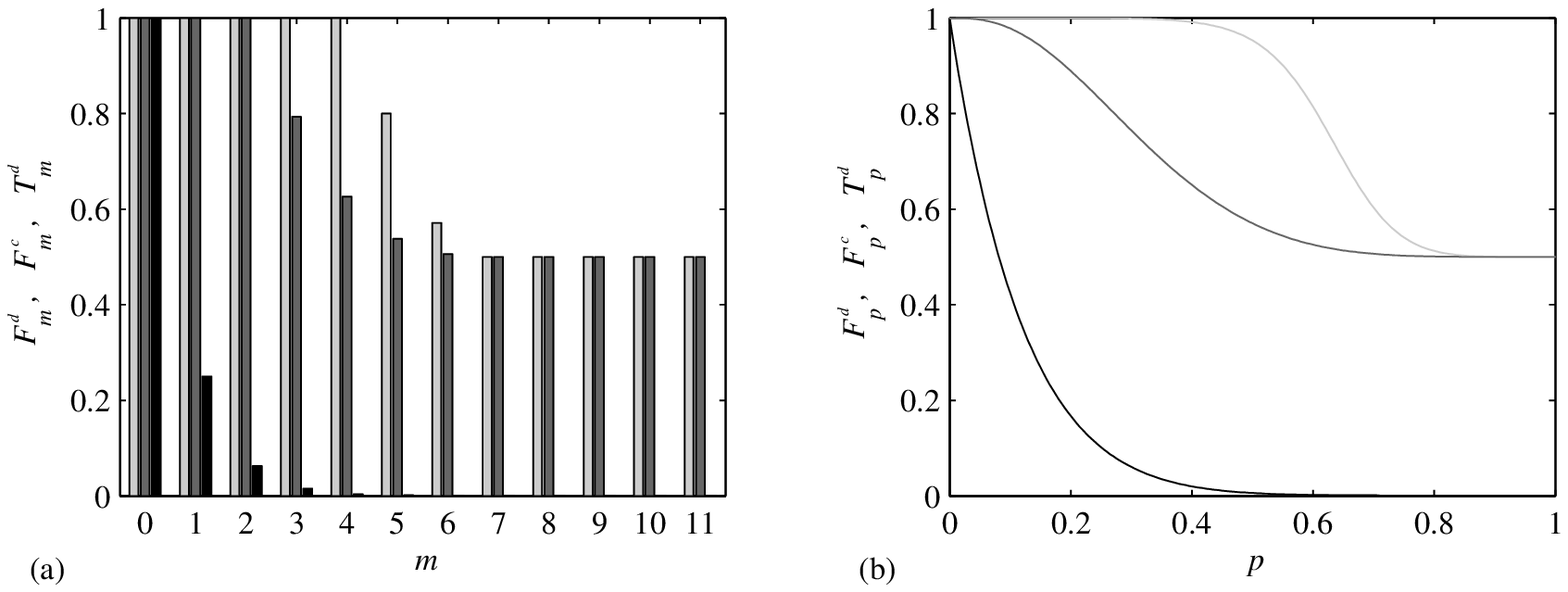}
\caption{For the $[[11,1,5]]$ code $\mathcal{G}_{11}$ we plot (a) from left to right, the transmission fidelity on $m$ 
qubits under error detection $\Fd_m$ [Eq.~(\ref{thmF2})] (light), the transmission fidelity on $m$ qubits under error correction $\Fc_m$ [Eq.~(\ref{prop1})] (dark), the transmission rate 
on $m$ qubits under error detection $\T_m$ [Eq.~(\ref{thmT2})] (black), and (b) from top to bottom, $\Fd_p$ [Eq.~(\ref{thmF3})] (light), $\Fc_p$ [Eq.~(\ref{prop2})] (dark), and $\T_p$ [Eq.~(\ref{thmT3})] (black) for the depolarizing channel.} 
\label{fig5}
\end{figure}
\begin{figure}[t]
\includegraphics[scale=1]{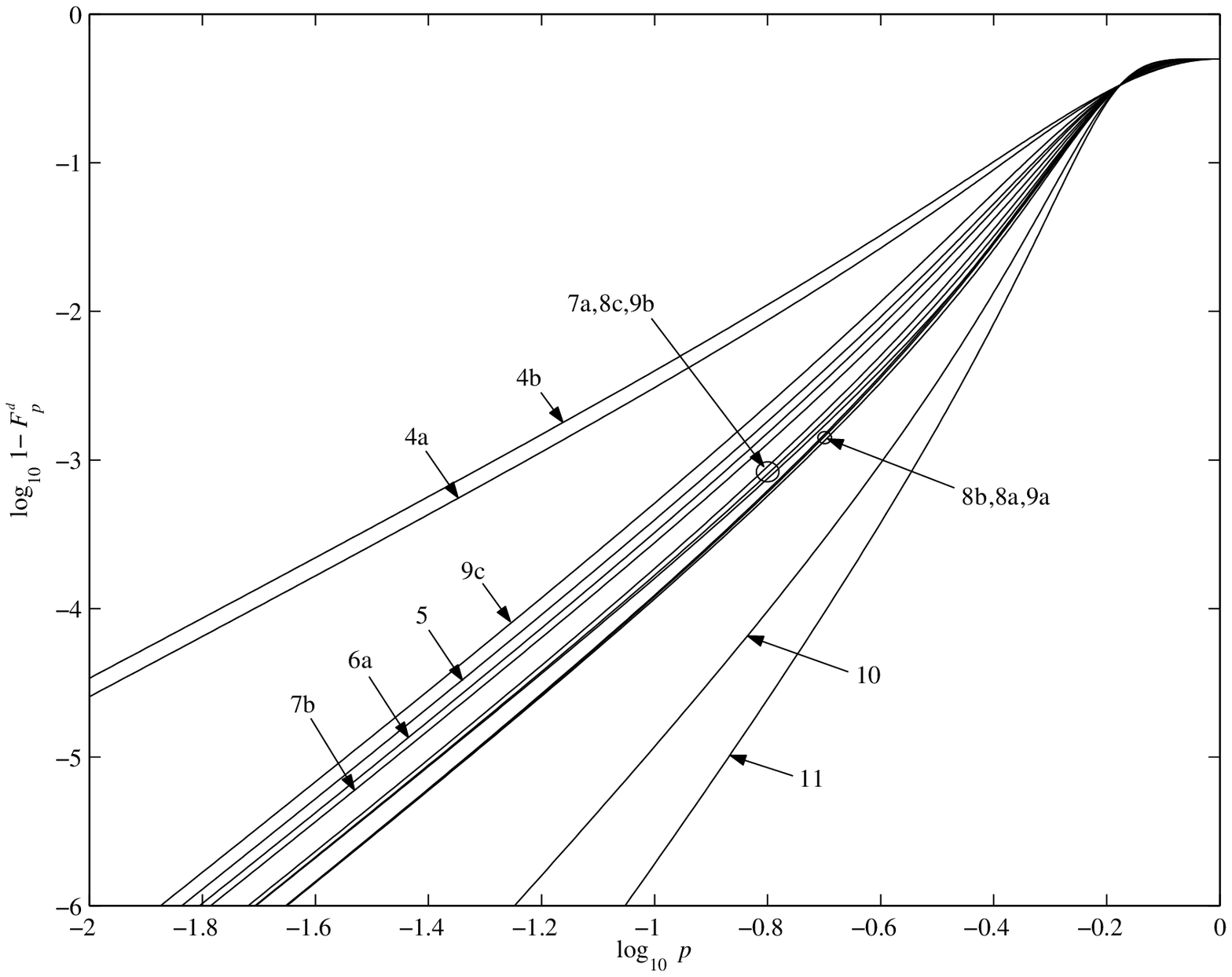}
\caption{The transmission fidelity for the depolarizing channel under error detection $\Fd_p$ [Eq.~(\ref{thmF3})].} 
\label{fig6}
\end{figure}
\begin{figure}[t]
\includegraphics[scale=1]{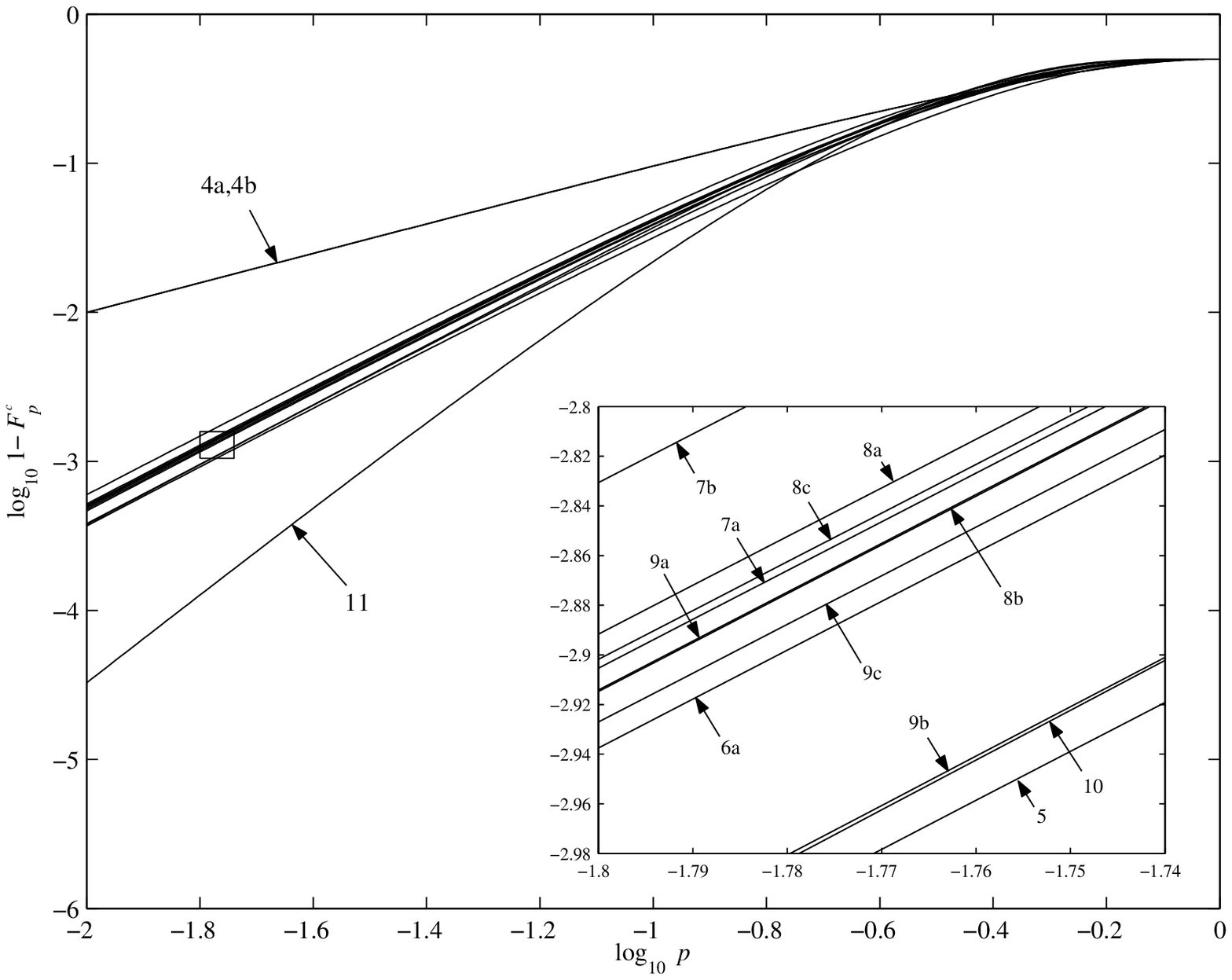}
\caption{The transmission fidelity for the depolarizing channel under error correction $\Fc_p$ [Eq.~(\ref{prop2})].} 
\label{fig7}
\end{figure}

Unlike in the previous cases, an exhaustive search over all correction operators $C_\lambda\in\mathcal{E}_\lambda$ is required to calculate the 
transmission fidelities $\Fc_m(P)$ and $\Fc_p(P)$. Although it is not apparent from Eq. (\ref{prop1}), given an $[[n,k,d]]_D$ 
code, $\Fc_m(P)=1$ for all $0\leq m <d/2$. In these cases we can choose $C_{\lambda}\in\mathcal{E}_{\lambda}$ to be of lowest 
weight in the set, and constant for all $0\leq m <d/2$. However, when $m\geq d/2$ (or when $p$ is large for $\Fc_p$) there may be more optimal 
choices. One example is the code generated by $\mathcal{G}_{7a}$ in the following section. Note from Eq.'s (\ref{Plambda}) and (\ref{extASp}) 
that for stabilizer codes
\begin{eqnarray}
A_S'(FP) &=& \frac{1}{D^{|S|}D'}\left|\left\{E\in F\mathcal{S}\,\big|\,\supp E\subseteq S\right\}\right| \\
A_m(FP) &=& \frac{1}{D'}\left|\left\{E\in F\mathcal{S}\,\big|\,\wt E=S\right\}\right|
\end{eqnarray}
when $F\in\mathcal{E}$, and from either of which we can calculate $A_m'(FP)$.

We may also define the probability of failure under error correction. The 
{\it failure rate on qudits $S$ under error correction}, 
\begin{equation}
\fc_S(P)\equiv 1-\Fc_S(P) =\frac{K\left[B_S'(P)-A_S'(P)\right]}{(K+1)B_S'(P)},
\end{equation}
and similarly, the {\it failure rate on $m$ qudits under error correction} is $\fc_m(P)\equiv 1-\Fc_m(P)$, and the 
{\it failure rate for the depolarizing channel under error correction} is $\fc_p(P)\equiv 1-\Fc_p(P)$.

Finally, to investigate the fidelity when $p$ is small, we set each $C_{\lambda}=C_{\lambda}(p)$, the optimal correction 
operator used for the maximization in Eq. (\ref{prop2}). Now given that $\Fc_m(P)=1$ for all $0\leq m< d'\equiv\lceil d/2\rceil$, 
when $p$ is small enough the operators $C_{\lambda}(p)$ may also be used for the maximizations in Eq. (\ref{prop1}) ($0\leq m< d'$). 
Thus, we must have 
\begin{equation}
\sum_{\lambda}A_m'\left(C_{\lambda}(p)P\right)=\frac{n!}{m!(n-m)!D^n}
\end{equation}
for all $0\leq m< d'$, when $p$ is small enough. Setting $C_{\lambda}'=\lim_{p\rightarrow 0}C_{\lambda}(p)$, we have
\begin{eqnarray}
\Fc_p(P) &=& 1-\frac{K}{K+1}\left\{1-\sum_{m=0}^n p^m(1-p)^{n-m} D^{-m}\sum_{\lambda}A_m'\left(C_{\lambda}'P\right)\right\} \\
&=& 1-\frac{K}{K+1}\left\{\sum_{m=0}^n p^m(1-p)^{n-m}\left[\frac{n!}{m!(n-m)!}-D^{-m}\sum_{\lambda}A_m'\left(C_{\lambda}'P\right)\right]\right\} \\
&=& 1-\frac{K}{K+1}\left\{\sum_{m=m'}^n p^m(1-p)^{n-m}\left[\frac{n!}{m!(n-m)!}-D^{-m}\sum_{\lambda}A_m'\left(C_{\lambda}'P\right)\right]\right\} \\
&=& 1-\frac{K}{K+1}\left[\frac{n!}{d'!(n-d')!}-D^{-d'}\sum_{\lambda}A_{d'}'\left(C_{\lambda}'P\right)\right]p^{d'}+O\left(p^{d'+1}\right) \label{correctsmallp}
\end{eqnarray}
in the limit $p\rightarrow 0$. Consequently, for error correction, we define 
\begin{equation}
c' \equiv\frac{K}{K+1}\left[\frac{n!}{d'!(n-d')!}-D^{-d'}\sum_{\lambda}A_{d'}'\left(C_{\lambda}'P\right)\right]. \label{correctsmallpc}
\end{equation}
If the operators $C_{\lambda}'$ can also be used for the maximization in Eq. (\ref{prop1}) when $m=d'$, then $c'=n!\fc_{d'}(P)/d!(n-d)!$.  
In all examples in the following section this was the case.

\section{Examples of stabilizer codes for qubits}
\label{sec5}

A classical {\it additive code over $GF(4)$ of length $n$} is an additive subgroup $\mathcal{C}$ of $GF(4)^n$.
In the case of qubits, stabilizer codes correspond to classical additive codes over $GF(4)$ \cite{calderbank}. 
This is shown as follows. Letting $GF(4)=\{0,1,\w,\wb\}$ where 
$\wb=\w^2=1+\w$, we define the {\it conjugate} of $x\in GF(4)$, denoted $\overline{x}$, by the mapping 
$\overline{0}=0$, $\overline{1}=1$, and $\overline{\wb}=\w$. Next define the {\it trace} map 
$\Tr:GF(4)\rightarrow GF(2)$ by $\Tr(x)=x+x^2$ i.e. $\Tr(0)=\Tr(1)=0$ and $\Tr(\w)=\Tr(\wb)=1$, 
and the {\it trace inner product} of two vectors ${\bf x} = x_1\dots x_n$ and ${\bf y} = y_1\dots y_n$ in $GF(4)^n$
as
\begin{equation}
{\bf x\star y} = \sum_{i=1}^n\Tr\big(x_i\overline{y_i}\big) \;\;\in GF(2).
\end{equation} 
The {\it weight} $\wt({\bf x})$ of ${\bf x}\in GF(4)^n$ is the number of nonzero 
components of ${\bf x}$, and the {\it minimum weight} of a code $\mathcal{C}$ is the smallest 
weight of any nonzero codeword in $\mathcal{C}$. Next, by defining the mapping $\Phi:GF(4)^n\rightarrow\mathcal{E}$ by $\Phi({\bf x})=\mathcal{D}(\phi^{-1}({\bf x}))$ where 
$\phi(\bm{\mu},\bm{\nu})=\w\bm{\mu}+\wb\bm{\nu}$, we can associate elements of $GF(4)$ with Pauli matrices 
($\w\rightarrow X$, $\wb\rightarrow Z$, $1\rightarrow iXZ$, $0\rightarrow I$), 
addition of vectors over $GF(4)^n$ with multiplication of operators in $\mathcal{E}$ 
(neglecting phases), and the trace inner product on $GF(4)^n$ with the commutator on 
$\mathcal{E}$. 

If $\mathcal{C}$ is an additive code, its {\it dual} is the additive code 
$\mathcal{C}^\perp=\{{\bf x}\in GF(4)^n \,|\, {\bf x\star c}=0 \;\forall\, {\bf c}\in\mathcal{C}\}$.
The code $\mathcal{C}$ is called {\it self-orthogonal} if $\mathcal{C}\subseteq\mathcal{C}^\perp$ and 
{\it self-dual} if $\mathcal{C}=\mathcal{C}^\perp$. The following theorem now applies \cite{calderbank}:
Suppose $\mathcal{C}$ is a self-orthogonal additive subgroup of $GF(4)^n$, containing $2^{n-k}$ vectors, such that 
there are no vectors of weight $<d$ in $\mathcal{C}^\perp\backslash\mathcal{C}$. Then any joint eigenspace of 
$\Phi(\mathcal{C})$ is an $[[n,k,d]]$ quantum error-correcting code. 

We say that $\mathcal{C}$ is {\it pure} if there are no nonzero vectors of weight $<d$ in $\mathcal{C}^\perp$. 
The associated quantum code is then pure if and only if $\mathcal{C}$ is pure. By convention, an $[[n,0,d]]$ code
corresponds to a self-dual additive code $\mathcal{C}$ with minimum weight $d$. Such codes are thus always pure. 

The {\it weight enumerators} of an additive code $\mathcal{C}$ are the Shor-Laflamme weight 
enumerators [Eq.'s (\ref{Ashor}) and (\ref{Bshor})] for the corresponding quantum code
\begin{eqnarray}
A_m(P) &=& A_m(\mathcal{C}) \equiv |\{{\bf x}\in\mathcal{C}\,|\,\wt({\bf x})=m\}|\\
B_m(P) &=& B_m(\mathcal{C}) \equiv |\{{\bf x}\in\mathcal{C^\perp}\,|\,\wt({\bf x})=m\}|.
\end{eqnarray}
The Rains enumerators may then be found through Eq.'s (\ref{rel1}) and (\ref{rel2}), or through 
their definition [Eq.'s (\ref{Amp}) and (\ref{Bmp})] with   
\begin{eqnarray}
A_S'(P) &=& A_S'(\mathcal{C}) \equiv \frac{1}{2^{|S|}}|\{{\bf x}\in\mathcal{C}\,|\,\supp({\bf x})\subseteq S\}|\\
B_S'(P) &=& B_S'(\mathcal{C}) \equiv \frac{1}{2^{|S|}}|\{{\bf x}\in\mathcal{C^\perp}\,|\,\supp({\bf x})\subseteq S\}|
\end{eqnarray}
where $\supp({\bf x})$ is the subset of $\{1,\dots,n\}$ consisting of all indices labeling a 
nonzero component of ${\bf x}$. Finally, note that when $\Phi({\bf y})=F$ we have
\begin{eqnarray}
A_S'(FP) &=& A_S'({\bf y}+\mathcal{C}) \equiv \frac{1}{2^{|S|}}|\{{\bf x}\in{\bf y}+\mathcal{C}\,|\,\supp({\bf x})\subseteq S\}| \\
A_m(FP) &=& A_m({\bf y}+\mathcal{C}) \equiv |\{{\bf x}\in{\bf y}+\mathcal{C}\,|\,\wt({\bf x})\subseteq S\}|.
\end{eqnarray}

The advantage of making the above correspondence is that a wealth of classical 
coding theory immediately becomes available. Indeed the classical Hamming code with 
generator matrix
\begin{equation}
\mathcal{G}_5=
\left[\begin{array}{ccccc}
\w&\wb&\wb&\w&0 \\        
0&\w&\wb&\wb&\w \\        
\w&0&\w&\wb&\wb \\        
\wb&\w&0&\w&\wb           
\end{array}\right] \label{c513}
\end{equation}
gives the quantum $[[5,1,3]]$ code. The rows of the generator matrix define 
a basis (under addition) for the classical code $\mathcal{C}$, and, with the above correspondence, define generators 
(up to a phase) for the stabilizer $\mathcal{S}$ in the quantum version. 

Two additive codes $\mathcal{C}_1$ and $\mathcal{C}_2$ are said to be {\it equivalent} when there exists a map between 
codewords of $\mathcal{C}_1$ and codewords of $\mathcal{C}_2$ consisting of a permutation of coordinates, 
a scaling of coordinates by elements of $GF(4)$, and conjugation of some of the coordinates. The quantum Hamming code 
above is unique under equivalence. We now catalogue other inequivalent additive codes whose quantum analogues encode a 
single qubit i.e. $k=1$. 

Exhaustive searches show that 
\begin{equation}
\mathcal{G}_{4a}=
\left[\begin{array}{cccc}
0&\wb&\w&\wb \\          
1&\w&\w&1 \\             
\w&\wb&\wb&\w            
\end{array}\right] ,
\qquad      
\mathcal{G}_{4b}=
\left[\begin{array}{cccc}
\w&\w&\w&\w \\           
\wb&\wb&0&0 \\           
0&0&\wb&\wb              
\end{array}\right]  ,     
\end{equation}
and
\begin{equation}
\mathcal{G}_{6a}=
\left[\begin{array}{cccccc}
0&0&0&0&1&1 \\             
0&1&1&1&1&0 \\             
0&\w&\w&\w&\w&\w \\        
1&0&1&\w&\wb&\w \\         
\w&0&\w&\wb&1&0            
\end{array}\right]   ,      
\qquad      
\mathcal{G}_{6b}=1\oplus\mathcal{G}_5  \:,    
\end{equation}
generate, respectively, the only inequivalent $[[4,1,2]]$ and $[[6,1,3]]$ codes \cite{calderbank}. The 4-qubit codes are pure while the 
6-qubit codes are necessarily impure. The code generated by $\mathcal{G}_{6b}$ offers no advantage over the 5-qubit code and will not be 
investigated any further.

Rows of the matrices
\begin{equation}
\mathcal{G}_{7a}=
\left[\begin{array}{ccccccc}
0&0&0&1&1&1&1 \\            
0&0&1&0&\w&\wb&1 \\         
0&1&0&\w&\wb&\wb&\w \\      
0&\w&\w&\w&1&\wb&0 \\       
1&0&0&0&\wb&\w&1 \\         
\w&0&\w&\w&\wb&\w&\wb       
\end{array}\right]  ,        
\qquad      
\mathcal{G}_{7b}=
\left[\begin{array}{ccccccc}
0&0&0&\w&\w&\w&\w \\        
0&\w&\w&0&0&\w&\w \\        
\w&0&\w&0&\w&0&\w \\        
0&0&0&\wb&\wb&\wb&\wb \\    
0&\wb&\wb&0&0&\wb&\wb \\    
\wb&0&\wb&0&\wb&0&\wb       
\end{array}\right]  ,
\end{equation}
generate inequivalent $[[7,1,3]]$ codes. These may be found by puncturing (Theorem 6b of \cite{calderbank}) the extremal self-dual additive codes of length 8 
found by Gaborit {\it et al} \cite{gaborit}. Both codes are pure. The code generated by $\mathcal{G}_{7b}$ 
is the Steane code \cite{steane}. More inequivalent $[[7,1,3]]$ codes will exist. For example, we can always trivially extend 
lower dimensional codes as done in the case of $\mathcal{G}_{6b}$.

Examples of $[[8,1,3]]$ codes follow from the matrices
\begin{equation}
\mathcal{G}_{8a}=
\left[\begin{array}{cccccccc}
0&0&0&1&1&1&1&1 \\           
0&0&0&\w&\w&\w&\w&0 \\       
0&0&1&0&1&\w&1&\wb \\        
0&1&0&0&\w&\wb&\wb&\w \\     
0&\w&\w&0&1&0&1&1 \\         
1&0&0&0&0&1&\wb&\wb \\       
\w&0&\w&0&\wb&1&0&\w         
\end{array}\right]      ,     
\qquad      
\mathcal{G}_{8b}=
\left[\begin{array}{cccccccc}
0&0&0&1&1&\wb&\w&0 \\        
0&0&1&0&\w&\w&0&1 \\         
0&0&\w&\w&0&\w&0&\w \\       
0&1&0&0&0&1&\wb&\wb \\       
0&\w&0&\w&\w&1&1&\wb \\      
1&0&0&0&\wb&0&1&\w \\        
\w&0&0&\w&0&\w&\w&0          
\end{array}\right]     ,      
\end{equation}
\begin{equation}
\mathcal{G}_{8c}=
\left[\begin{array}{cccccccc}
0&0&0&1&1&1&1&1 \\           
0&0&0&\w&\w&\w&\w&0 \\       
0&0&1&0&1&\wb&\w&0 \\        
0&1&0&0&\w&\w&0&1 \\         
0&\w&\w&0&0&\w&0&\w \\       
1&0&0&0&0&1&\wb&\wb \\       
\w&0&\w&0&\w&1&1&\wb         
\end{array}\right] .          
\end{equation}
All of these codes are pure and were found by puncturing the extremal self-dual additive codes of length 9 in \cite{gaborit}.
Again, more inequivalent $[[8,1,3]]$ codes will exist.

Examples of $[[9,1,3]]$ codes include
\begin{equation}
\mathcal{G}_{9a}=
\left[\begin{array}{ccccccccc}
0&0&0&0&\w&\wb&\wb&0&\wb \\   
0&0&0&1&0&1&0&1&1 \\          
0&0&1&0&1&0&0&1&1 \\          
0&0&\w&\w&0&0&\wb&\wb&0 \\    
0&1&0&0&1&0&1&1&0 \\          
0&\w&0&\w&0&\wb&\wb&0&0 \\    
1&0&0&0&0&1&1&1&0 \\          
\w&0&0&\w&1&\w&\w&\w&\w       
\end{array}\right]    ,        
\qquad      
\mathcal{G}_{9b}=
\left[\begin{array}{ccccccccc}
0&0&0&0&\w&0&1&1&\w \\        
0&0&0&1&1&1&1&1&1 \\          
0&0&0&\w&0&\w&\wb&\wb&0 \\    
0&0&1&0&0&1&\w&0&\wb \\       
0&1&0&0&0&1&0&\w&\wb \\       
0&\w&\w&0&0&\wb&1&\wb&1 \\    
1&0&0&0&1&1&1&\wb&\w \\       
\w&0&\w&0&1&\wb&0&1&\wb       
\end{array}\right]   ,         
\end{equation}
\begin{equation}
\mathcal{G}_{9c}=
\left[\begin{array}{ccccccccc}
\wb&\wb&0&0&0&0&0&0&0 \\      
0&\wb&\wb&0&0&0&0&0&0 \\      
0&0&0&\wb&\wb&0&0&0&0 \\      
0&0&0&0&\wb&\wb&0&0&0 \\      
0&0&0&0&0&0&\wb&\wb&0 \\      
0&0&0&0&0&0&0&\wb&\wb \\      
\w&\w&\w&\w&\w&\w&0&0&0 \\    
0&0&0&\w&\w&\w&\w&\w&\w       
\end{array}\right] .          
\end{equation}
The first two codes, $\mathcal{G}_{9a}$ and $\mathcal{G}_{9b}$, are pure and were found by puncturing the extremal self-dual 
additive codes of length 10 found by Bachoc and Gaborit \cite{bachoc}. The code generated by $\mathcal{G}_{9c}$ 
is the impure Shor code \cite{shor2}. Many more inequivalent $[[9,1,3]]$ codes will exist.

Finally we give generator matrices for pure $[[10,1,4]]$ and $[[11,1,5]]$ codes, found by puncturing, 
respectively, the shortened dodecacode and dodecacode:
\begin{equation}
\mathcal{G}_{10}=
\left[\begin{array}{cccccccccc}  
\w&0&0&0&0&\wb&\w&\w&\wb&0 \\    
0&\w&0&0&\wb&1&\w&\wb&\wb&\w \\  
0&0&\w&0&0&\wb&1&\wb&\wb&\w \\   
0&0&0&\w&\wb&1&1&1&\w&1 \\       
0&0&0&0&\w&\wb&0&\wb&1&1 \\      
0&\wb&0&0&0&\w&1&1&1&1 \\        
0&0&\wb&0&\wb&0&\w&\w&\wb&\wb \\ 
\wb&\wb&\wb&0&\wb&0&\wb&\w&0&1 \\
\wb&0&0&\wb&0&0&\wb&\wb&\w&\w    
\end{array}\right] ,              
\qquad      
\mathcal{G}_{11}=
\left[\begin{array}{ccccccccccc}      
\w&0&0&0&0&\wb&\wb&\wb&0&\w&\w \\     
0&\w&0&0&0&\wb&\wb&1&\wb&0&1 \\       
0&0&\w&0&0&\wb&\wb&\w&\w&\wb&0 \\     
0&0&0&\w&0&\wb&1&0&\w&\w&1 \\         
0&0&0&0&\w&\wb&\w&0&\wb&\wb&\w \\     
0&0&\wb&0&0&\w&1&0&1&\wb&\wb \\       
0&\wb&\wb&0&0&0&\w&\wb&0&\w&\wb \\    
0&\wb&0&0&\wb&0&\wb&\w&\w&0&\wb \\    
\wb&\wb&\wb&0&\wb&0&0&\wb&\w&\wb&\w \\
0&\wb&0&\wb&0&0&\wb&0&\wb&\w&\w       
\end{array}\right]  .                  
\end{equation}

In Fig.'s \ref{fig1} through \ref{fig5} we plot the quantities $\T_m$, $\Fd_m$, $\Fc_m$, $\T_p$, $\Fd_p$, and $\Fc_p$ for 
the $[[4,1,2]]$ code $\mathcal{G}_{4a}$, the unique $[[5,1,3]]$ code $\mathcal{G}_{5}$, Stean's $[[7,1,3]]$ code $\mathcal{G}_{7b}$, 
Shor's impure $[[9,1,3]]$ code $\mathcal{G}_{9c}$, and the $[[11,1,5]]$ code $\mathcal{G}_{11}$. Next, in Fig.'s \ref{fig6} and \ref{fig7}, 
we plot $\log_{10} 1-\Fd_p$ and $\log_{10} 1-\Fc_p$, respectively, versus $\log_{10} p$ for all of the stabilizer codes given above. 

When $p$ is small the transmission fidelity under error detection follows from Eq. (\ref{detectsmallp}), and thus we can rank different 
codes using the pair $(d,c)$, where $c$ is defined in Eq. (\ref{detectsmallpc}). The codes in order are: 
$\mathcal{G}_{4b}$ (2,1/3), $\mathcal{G}_{4a}$ (2,1/4), $\mathcal{G}_{9c}$ (3,13/32), $\mathcal{G}_{5}$ (3,5/16), 
$\mathcal{G}_{6a}$ (3,1/4), $\mathcal{G}_{7b}$ (3,7/32), $\mathcal{G}_{7a}$ (3,13/96),$\mathcal{G}_{8c}$ (3,1/8), 
$\mathcal{G}_{9b}$ (3,1/8), $\mathcal{G}_{8b}$ (3,1/12), $\mathcal{G}_{8a}$ (3,1/12),$\mathcal{G}_{9a}$ (3,1/12), 
$\mathcal{G}_{10}$ (4,5/64), and $\mathcal{G}_{11}$ (5,33/256). 
Note that as the number of qubits increases we are generally able to construct better codes even when the minimum distance remains 
constant. However this is not the case for error correction. 

In the case of error correction we use the pair $(d',c')$ [see Eq.'s (\ref{correctsmallp}) and (\ref{correctsmallpc})]
to rank different codes. The codes in order are now:  
$\mathcal{G}_{4b}$ (1,1), $\mathcal{G}_{4a}$ (1,1), $\mathcal{G}_{7b}$ (2,49/8), $\mathcal{G}_{8a}$ (2,127/24), 
$\mathcal{G}_{8c}$ (2,31/6), $\mathcal{G}_{7a}$ (2,41/8), $\mathcal{G}_{9a}$ (2,5), $\mathcal{G}_{8b}$ (2,5), 
$\mathcal{G}_{9c}$ (2,39/8), $\mathcal{G}_{6a}$ (2,19/4), $\mathcal{G}_{9b}$ (2,23/6), $\mathcal{G}_{10}$ (2,15/4), 
$\mathcal{G}_{5}$ (2,15/4), and $\mathcal{G}_{11}$ (3,273/8). 
Note that the 5-qubit code outperforms all other codes of minimum distance 3. The 10-qubit code asymptotes to the five only 
at much smaller values of $p$ than shown in the inset of Fig. \ref{fig7}. Thus, for the codes investigated, the benefit of 
accessing more qubits to construct a code is outweighed by the cost of allowing the extra qubits into an error-prone environment.

\section{Conclusion}
\label{sec6}

In conclusion, we have investigated the performance of a quantum error-correcting code when stretched beyond its intended capabilities.
The content of Theorem's \ref{thmT}, \ref{thmF} and \ref{thm17}, along with Proposition \ref{prop} form the main results of the 
paper. We have derived the transmission rate, $\T$ (the probability that no error is detected), and the transmission fidelity, 
$\Fd$, under error detection, in Theorem's \ref{thmT} and \ref{thmF} respectively. In the error detection scenario a corrupted state 
is simply discarded once detected. Theorem \ref{thm17} and Proposition \ref{prop} are concerned with error correction. Here 
we attempt to correct all corrupted states. In this case we give expressions for transmission fidelity, $\Fc$, for stabilizer 
codes. The quantities $\Fc$, $\Fd$ and $\T$ in their various forms, or $c$ and $c'$ [Eq.s (\ref{detectsmallpc}) and (\ref{correctsmallpc})],
might be used to compare different quantum error-correcting codes of the same minimum distance. Indeed, under the depolarizing channel, the unique 
five-qubit quantum Hamming code outperforms other known codes of the same minimum distance in the error correction scenario,
but loses out to codes constructed from higher numbers of qubits in the error detection scenario.

\begin{acknowledgments}
The author would like to thank Bryan Eastin for helpful discussions. This work was supported in part 
by ONR Grant No.~N00014-00-1-0578 and by ARO Grant No.~DAAD19-01-1-0648.
\end{acknowledgments}

\end{document}